\documentclass[lettersize,journal]{IEEEtran}
\usepackage{amsmath,amsfonts}
\usepackage{algorithmic}
\usepackage{array}
\usepackage[caption=false,font=normalsize,labelfont=sf,textfont=sf]{subfig}
\usepackage{textcomp}
\usepackage{stfloats}
\usepackage{url}
\usepackage{verbatim}
\usepackage{graphicx}
\usepackage{cite}
\usepackage[vlined,linesnumbered,ruled]{algorithm2e}
\usepackage{tikz}
 
\newcommand*\halfcirc[1][1ex]{%
  \begin{tikzpicture}
  \draw[fill] (0,0)-- (90:#1) arc (90:270:#1) -- cycle ;
  \draw (0,0) circle (#1);
  \end{tikzpicture}}
\newcommand*\fullcirc[1][1ex]{\tikz\fill (0,0) circle (#1);}

\usepackage{tikz}
\usepackage{multirow}
\usepackage{longtable}
\usepackage{tabularx}
\usepackage{xurl}
\usepackage{hyperref}
\usepackage{lipsum}
\usepackage{threeparttable}
\usepackage{soul}

\usepackage{CJKutf8}

\usepackage{tikz}
\newcommand*\circled[1]{\tikz[baseline=(char.base)]{%
            \node[shape=circle,draw,inner sep=1pt] (char) {#1};}}
\usepackage{bm}

\begin{document}
\begin{CJK*}{UTF8}{bkai}

\title{SAGA: Synthetic Audit Log Generation for APT Campaigns}

\author{Yi-Ting Huang, Ying-Ren Guo, Yu-Sheng Yang, Guo-Wei Wong, Yu-Zih Jheng, Yeali Sun, Jessemyn Modini, Timothy Lynar, and~Meng Chang Chen
\thanks{Y.\ Huang is with National Taiwan University of Technology and Science.\protect}
\thanks{Y.\ Guo and M.\ Chen are with Academia Sinica, Taiwan.\protect\\
 E-mail: mcc@iis.sinica.edu.tw}
\thanks{J.\ Modini and T.\ Lynar are with University of New South Wales,  Australia.\protect}
\thanks{Y.\ Yang, Y.\ Jheng, G.\ Wong and Y.\ Sun are with National Taiwan University, Taiwan.\protect}}

\IEEEoverridecommandlockouts 

\IEEEpubid{
\begin{minipage}{\textwidth}\ \\[40pt] \centering
  \copyright~2025 IEEE. Personal use of this material is permitted. Permission from IEEE must be obtained for all other uses, in any current or future media, including reprinting/republishing this material for advertising or promotional purposes, creating new collective works, for resale or redistribution to servers or lists, or reuse of any copyrighted component of this work in other works.
\end{minipage}
}

\maketitle

\begin{abstract} 
With the increasing sophistication of Advanced Persistent Threats (APTs), the demand for effective detection and mitigation strategies and methods has escalated. Program execution leaves traces in the system audit log, which can be analyzed to detect malicious activities. However, collecting and analyzing large volumes of audit logs over extended periods is challenging, further compounded by insufficient labeling that hinders their usability. Addressing these challenges, this paper introduces SAGA (Synthetic Audit log Generation for APT campaigns), a novel approach for generating find-grained labeled synthetic audit logs that mimic real-world system logs while embedding stealthy APT attacks. SAGA generates configurable audit logs for arbitrary duration, blending benign logs from normal operations with malicious logs based on the definitions the MITRE ATT\&CK framework. Malicious audit logs follow an APT lifecycle, incorporating various attack techniques at each stage. These synthetic logs can serve as benchmark datasets for training machine learning models and assessing diverse APT detection methods. To demonstrate the usefulness of synthetic audit logs, we ran established baselines of event-based technique hunting and APT campaign detection using various synthetic audit logs. In addition, we show that a deep learning model trained on synthetic audit logs can detect previously unseen techniques within audit logs.
\end{abstract}


\section{Introduction}
\IEEEPARstart{T}{he} emergence of Advanced Persistent Threats (APTs), such as Fancy Bear (APT28) and Lazarus Group Elfin (APT33), poses significant challenges to the cybersecurity community.
APT campaigns, typically orchestrated by advanced threat actors with economic or political motives,involve multiple stages,
including gaining initial access to a victim environment, remaining undetected for extended periods, exfiltrating sensitive data, and compromising system integrity. Despite progress in malware analysis~\cite{MAMBA,APILI}, effective detection and mitigation of APT attacks remains a critical priority. A key obstacle in APT threat studies is the lack of a widely recognized benchmark dataset~\cite{zepperle2022}, which is essential to develop AI/ML-based solutions and objectively evaluate methods. This challenge motivates the study of synthetic audit log generation in this work.

System audit logs, which record traces of program execution, play a crucial role in log analysis for threat detection. For example, MITRE ATT\&CK Evaluations~\cite{attack_eval} provide emulation plans that allow cybersecurity providers to assess their product’s ability to detect simulated cyber attacks through system logs. 
With the advancement of data-driven artificial intelligence methods, it has become increasingly effective in addressing certain cyber threats, such as attack analysis \cite{alsaheel2021atlas, ding2023airtag, cheng2023kairos}, which offer the potential for timely Technique detection.
These approaches typically rely on high-quality labeled audit logs for model training. 
In addition, it allows for the enhancement and comparison of various methods.

However, labeling large volumes of audit logs over long periods poses challenges, especially when APT attacks lack obvious indicators. Researchers often rely on publicly available datasets such as the DARPA Transparent Computing Engagement~\cite{darpa2tc}, Operationally Transparent Cyber (OpTC)\cite{darpa2optc}, and LANL’s dataset\cite{kent2015comprehensive}, as well as smaller datasets that are rarely shared due to privacy and proprietary concerns. 
Moreover, some datasets provide fine-grained labels on malicious entities~\cite{alsaheel2021atlas} or processes~\cite{liu2025we}, while others~\cite{darpa2tc,darpa2optc} only report attack scenarios over time intervals, without linking them to specific system events.
This shortage of labeled logs limits a deeper understanding of attack behaviors and restricts the development of AI/ML-based cybersecurity defense studies~\cite{arp2022and, jacobs2022ai}.

The absence of such audit logs motivates the development of methods for generating configurable synthetic audit logs that fulfill the following key requirements.
\begin{itemize}
    \item \textbf{Realism.} Synthetic audit log should closely resemble the real audit logs.
    \item \textbf{Scenario coverage.} They should represent a broad spectrum of APT campaigns, including recent and emerging threats.
    \item \textbf{Detailed labeling.} Logs should include fine-grained labels such as specific Techniques or equivalents mapped to events in the audit log.
    \item \textbf{Higher-Level representation.} Along with raw events, they should offer higher-level abstractions, facilitating easier understanding, examination, and communication.
    \item \textbf{Flexibility.} The framework should be able to generate logs of any duration and include any number of APT campaigns.
    \item \textbf{Diversity.} The framework should be able to generate previously unseen logs, which would benefit model training and examination.
\end{itemize}

This study introduces SAGA (Synthetic Audit Log Generation for APT Campaigns), a framework for generating configurable synthetic audit logs. SAGA is aligned with the MITRE ATT\&CK framework and terminologies, chosen for its openness, comprehensiveness, and availability. 
The generated synthetic audit log contains both benign and malicious (APT campaigns) audit logs, with benign logs representing normal system and user operations and malicious logs constructed from pre-defined Technique templates. Each APT campaign in the logs, comprising stages/tactics with corresponding techniques, which are reflected in the audit logs. 

The specific tasks of this study include:
\begin{itemize}
    \item \textbf{Obtaining attack scenarios.} Use a red team emulator to collect and label logs with technique identifiers based on the MITRE ATT\&CK framework. 
 \item \textbf{Designing the SAGA framework.} Develop the SAGA framework and workflow to address the aforementioned challenges. 
\item \textbf{Synthesizing configurable audit logs.} Create synthetic audit logs that combine APT attack scenarios and benign behaviors, resembling real data logs.
 \item \textbf{Verifying SAGA usefulness.} Run existing systems using the SAGA synthetic audit log as input to demonstrate its effectiveness. Furthermore, show that models trained on the synthetic audit log can successfully identify previously unseen malicious behaviors in new datasets.
\end{itemize}

To demonstrate the usefulness of synthetic audit logs, we tested established baselines and the existing methods across various applications, including APT intrusion detection, technique hunting, and APT campaign detection. The objective of this empirical study is to demonstrate that SAGA synthetic audit logs can be used effectively and meaningfully in cybersecurity-related audit log applications. The focus is not on comparing the performance or goodness of these methods, as they target different objectives and operate in distinct contexts, making direct comparison uninformative.

\section{Background and Motivation}
In this section, we introduce the audit log and its related issues using a running example of an APT campaign. 

\subsection{System Audit Log}

System audit logs are records of system events such as file access, process creation and termination, and network access. Each system event depicts the interaction between two system entities, typically represented as \textless{}subject, operation, object\textgreater{}, where the subject refers to a process entity or a program, and the object represents a system entity such as a process, file, registry, or network socket. The operation signifies the action performed by the subject on the object. 
For example, the event \textless{}\textit{WinRAR.exe}, \textit{CreateFile}, \textit{IOC\_09\_11.rar}\textgreater{} indicates that the \textit{WinRAR} process creates a file named \textit{IOC\_09\_11.rar}. 
System events, along with their subjects and objects, include attributes such as meta-information, properties, and other details provided by the system, which can further describe the event and context. These details will be discussed in the subsequent section.
System audit logs can be obtained using built-in kernel event logging mechanisms such as \textit{Windows Event Tracing (ETW)} and \textit{Linux Audit}, or through additional tools like \textit{Procmon}, \textit{Sysmon}, and \textit{CamFlow}.

\begin{figure}[!htb]
    \centering
    \includegraphics[width=0.48\textwidth]{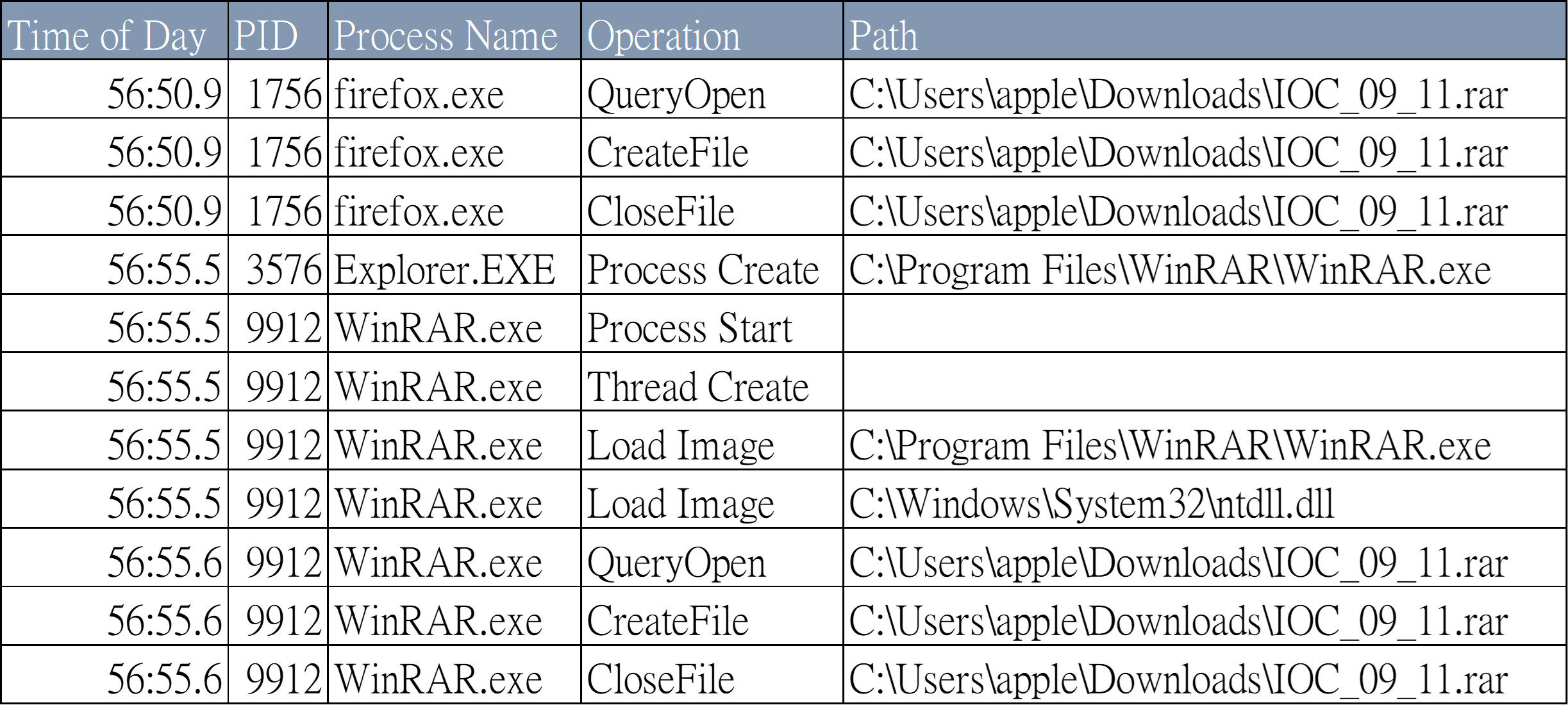}
    \caption{Example of audit log events captured by \textit{Procmon}.}
    \label{fig:audit_log}
\end{figure}

In this study, we use \textit{Procmon} as the primary logging tool. \textit{Procmon}, short for Process Monitor is a Windows utility that monitors and records real-time system activities, including process actions, file system activities, registry changes, and more. Figure~\ref{fig:audit_log}
provides an example of the captured system events. The first row records the event \textless{}\textit{firefox.exe}, \textit{QueryOpen}, \textit{...\textbackslash IOC\_09\_11.rar}\textgreater{}, along with attributes such as timestamp, PID, and object value, which indicate a query for the existence of the RAR file.
The audit log is proved useful for analyzing malicious attempts and APT attacks, offering insights into their characteristics.

\subsection{APT Attack Lifecycle}
\label{sec: lifecycle}

\begin{figure*}[!htb]
    \centering
    \includegraphics[width=0.9\textwidth]{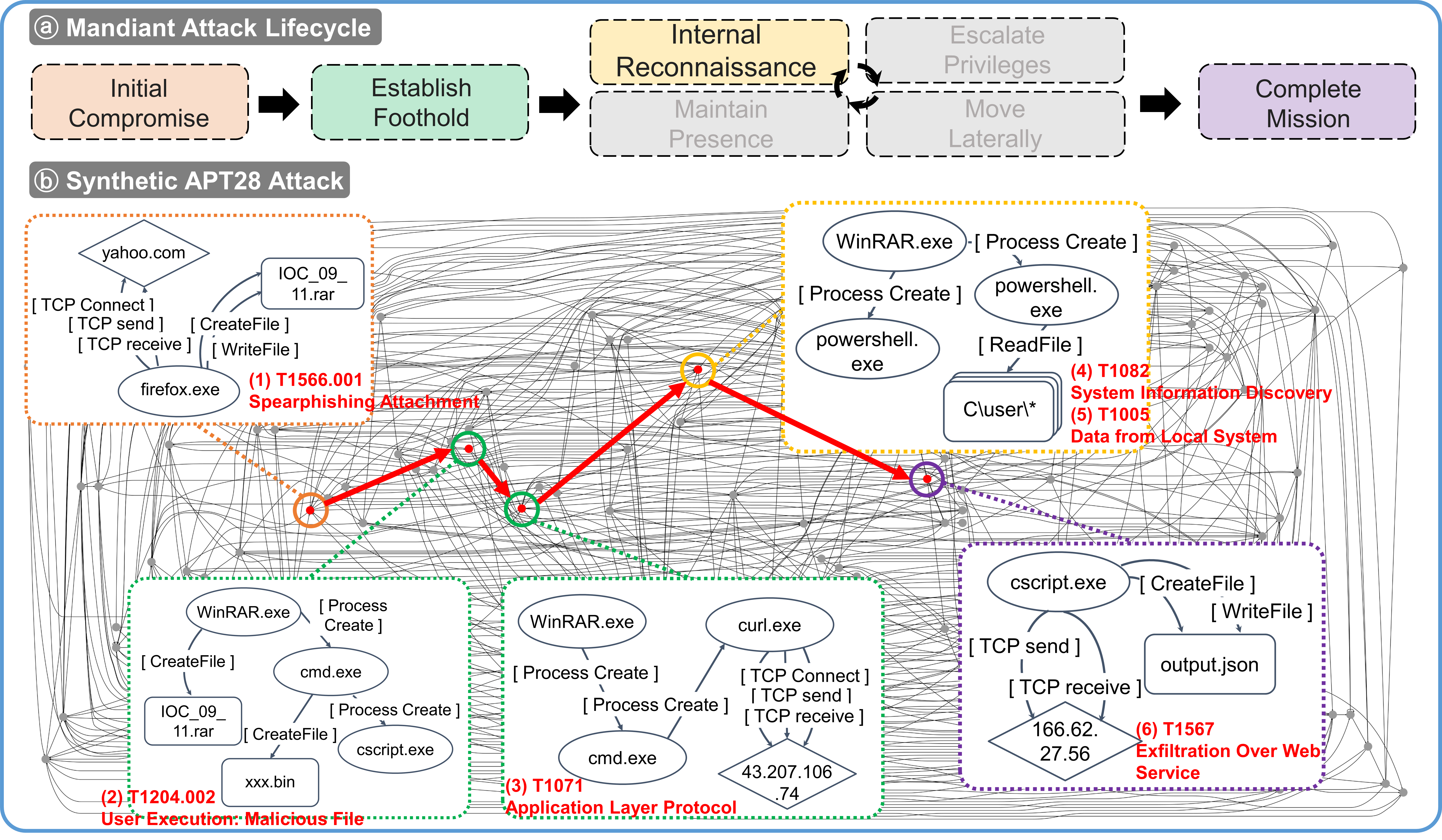}
    \caption{
        \textcircled{a} 
        Mandiant attack lifecycle~\cite{mandiantexposing} of APT28. 
        \textcircled{b} 
    The APT28 attack provenance graph, derived from~\cite{APT28}, presents five malicious events labeled with six techniques, which are mapped to the corresponding stages of the attack lifecycle. In each zoomed-in box, the dotted boundary color corresponds to its specific stage in the attack lifecycle. Rectangular nodes represent files, diamond-shaped nodes represent sockets, oval nodes denote processes, and the edges illustrate the causal relationships between the entities. The red arrow edges indicate the sequence of malicious events.
    } 

    \label{fig:Running_example}
\end{figure*}

Frameworks such as Lockheed Martin's cyber kill chain~\cite{martin2015gaining}, MITRE ATT\&CK framework~\cite{strom2018mitre}, and Mandiant adversary lifecycle~\cite{mandiantexposing} define the lifecycle of cyber attacks, helping defenders understand the collective operations and intention of APT attacks. This study uses the Mandiant adversary lifecycle, depicted in Figure~\ref{fig:Running_example} \circled{a}, as a reference for the APT attack lifecycle, although other frameworks can be equally applicable. Each stage/tactic encompasses associated Techniques, and each Technique may have several implementations/abilities. In addition, this lifecycle framework helps to generate and analyze synthetic audit logs. Note that in this paper, we consider Technique and Sub-Technique equally and use the term Technique to represent both.

A typical APT attack consists of multiple stages, including initial compromise, establishing foothold, internal reconnaissance, escalating privileges, moving laterally, maintaining presence, and ultimately complete mission. Note that the stages of internal reconnaissance, escalating privileges, moving laterally, and maintaining presence may occur multiple times, and not all four stages are necessary in every instance.
Figure~\ref{fig:Running_example} \circled{a} shows the lifecycle stages corresponding to the APT28 campaign.

\subsection{Running Example}
\label{sec: example}

We refer to APT28 threat intelligence reports~\cite{APT28} to construct the attack scenario illustrated in Figure~\ref{fig:Running_example}, which serves as the running example in this work. Figure~\ref{fig:Running_example} \circled{a} outlines its attack lifecycle, while Figure~\ref{fig:Running_example} \circled{b} shows the corresponding APT audit log. 
In the initial stages of the attack, phishing emails containing malicious attachments are sent to the victim to download the attachment \textit{IOC\_09\_11.rar}. This activity is designated as the Technique \textit{T1566.001 Spearphishing Attachment}, as (1) in Figure~\ref{fig:Running_example} \circled{b}, of the attack lifecycle stage \textit{Initial Compromise}. This malicious attachment exploits a vulnerability in WinRAR versions below 6.23 (as mentioned in CVE-2023-38831). 
When the victim clicks on the \textit{IOC\_09\_11.rar} file, it triggers subsequent malicious activities designated as \textit{ T1204.002 User Execution: Malicious File} 
of the attack lifecycle \textit{Establish Foothold}, as (2) in Figure~\ref{fig:Running_example} \circled{b}.

The attack proceeds by establishing a reverse shell between the attacker and the victim, designated as \textit{T1071 Application Layer Protocol} of the \textit{Establish Foothold} stage, marked as (3) in the figure. Next, it steals information such as local state files, corresponding to \textit{T1082 System Information Discovery} and \textit{T1005 Data from Local System}, marked as (4) and (5) of the \textit{Internal Reconnaissance} stage. Finally, the stolen information is uploaded to a web \textit{servicewebhook$.$site}, designated as \textit{T1567 Exfiltration Over Web Service} of the \textit{Complete Mission} stage, shown as (6) in Figure~\ref{fig:Running_example} \circled{b}.

\section{SAGA Generative Model}
\label{sec: generativeThreatModel}

\begin{figure}[htb!]
    \centering
    \includegraphics[width=0.48\textwidth]{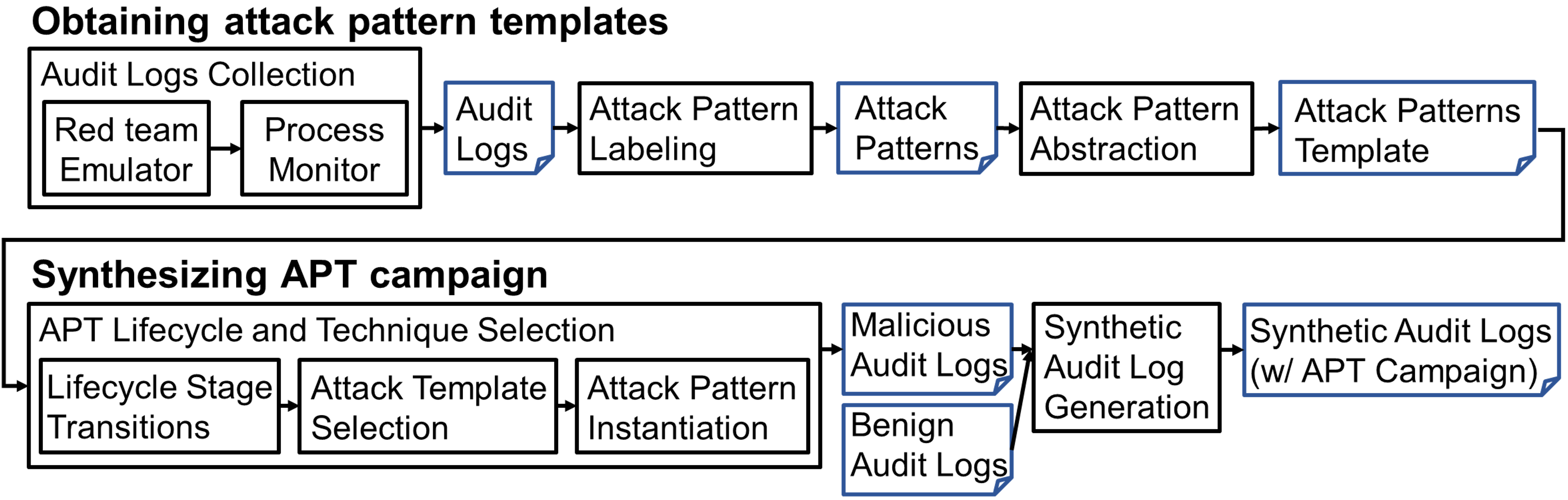}
    \caption{SAGA workflow}
    \label{fig:system_overview}
\end{figure}

The SAGA generative model involves collecting audit logs from a red-team emulator to create a repository of attack pattern templates with fine-grained labels. These templates are then used, following lifecycle selection instructions, to generate synthetic audit logs that embed APT campaigns. 
Figure~\ref{fig:system_overview} provides an overview and the workflow of SAGA, which will be discussed in detail in the following subsections.
To mitigate the risk of detection models overfitting on artificial patterns, SAGA introduces variability in both attack implementation in Subsection~\ref{sec: labeling} and artifact representation in Subsection~\ref{sec: substitution}.
In Subsection~\ref{sec: labeling}, SAGA collects attack pattern from a red-team emulation platform that supports multiple abilities per Technique, enabling the generation of diverse execution paths and behavioral variants.
In Subsection~\ref{sec: substitution}, SAGA incorporates entity-level variability by replacing generalized entities with hyponyms derived from real-world malware samples and synthetic variants.
This combination of implementation diversity and entity-level variation helps generate a wide range of realistic log instances.

\subsection{Attack Pattern Collection and Labeling}\label{sec: labeling}
The first task of SAGA is to collect labeled audit logs, as depicted in Figure~\ref{fig:system_overview}. 
The Red Team Platform, such as CALDERA~\cite{applebaum2016caldera,caldera} and Atomic Red Team~\cite{atomic}, simulates real-world adversaries in a controlled environment.
In this study, we use CALDERA to emulate real-world attacks by adversaries. Our choice is based on its open source nature and alignment with the MITRE ATT\&CK framework.
To deploy CALDERA, we configure a master server and set up Remote Access Tool (RAT) clients on a compromised host.
An \textit{ability} of CALDERA refers to a specific implementation of an ATT\&CK Technique, representing a discrete adversarial behavior. 
After acquiring audit logs from a red team emulator and process monitor, the next step is to extract the relevant attack patterns and assign appropriate Technique labels. CALDERA-generated reports provide details for each executed attack, including process IDs, commands used, and associated technique IDs. 
We employ a semi-automated labeling method to annotate audit logs based on handcrafted rules tailored to specific attack techniques.

To develop specific rules for each ability, three authors applied their expert knowledge based on the MITRE ATT\&CK framework~\cite{strom2018mitre} to define rules for identifying attack-related events.
In the case of T1566.001, the ability of T1566.001 involves using the \textit{Invoke-WebRequest} cmdlet of PowerShell, which indicates network activity connecting to the GitHub website. Furthermore, the flag \textit{-OutFile} and parameter \textit{\textbackslash\$env:TEMP\textbackslash PhishingAttachment.xlsm} suggests that a file named \textit{PhishingAttachment.xlsm} will be created on the target system. 
To ensure accurate labeling, rules are designed and applied using both forward and backward tracing, since not all events generated during an attack should be considered malicious and labeled with the Technique identifier. Some may represent background or system activities. 
Only events exhibiting these specific malicious actions and artifacts are considered malicious for this particular ability, and custom labeling rules are crafted to capture this behavior. 
We validated these rules using a set of manually labeled logs and through cross-verification via code review among the three authors.

When given a CALDERA-generated report, SAGA uses the attack-related Process ID (PID) as the root to construct a comprehensive process family tree that captures all associated processes.
By mapping these process IDs to the corresponding audit logs and applying handcrafted rules, SAGA can accurately assign the appropriate technique ID to the relevant log entries.

\subsection{Attack Pattern Templates Abstraction} \label{sec: template}
Each attack pattern corresponds to a specific instance of a Technique with particular parameter values. To enable the generation of a wider variety of attack patterns, it is essential to abstract these patterns into a more abstract form. These abstractions can then be instantiated with legitimate parameter values, allowing for the creation of new attack patterns. This approach facilitates greater flexibility and diversity in the generation of new attack instances.

\begin{figure}[tb]
    \centering
    \includegraphics[width=0.48\textwidth]{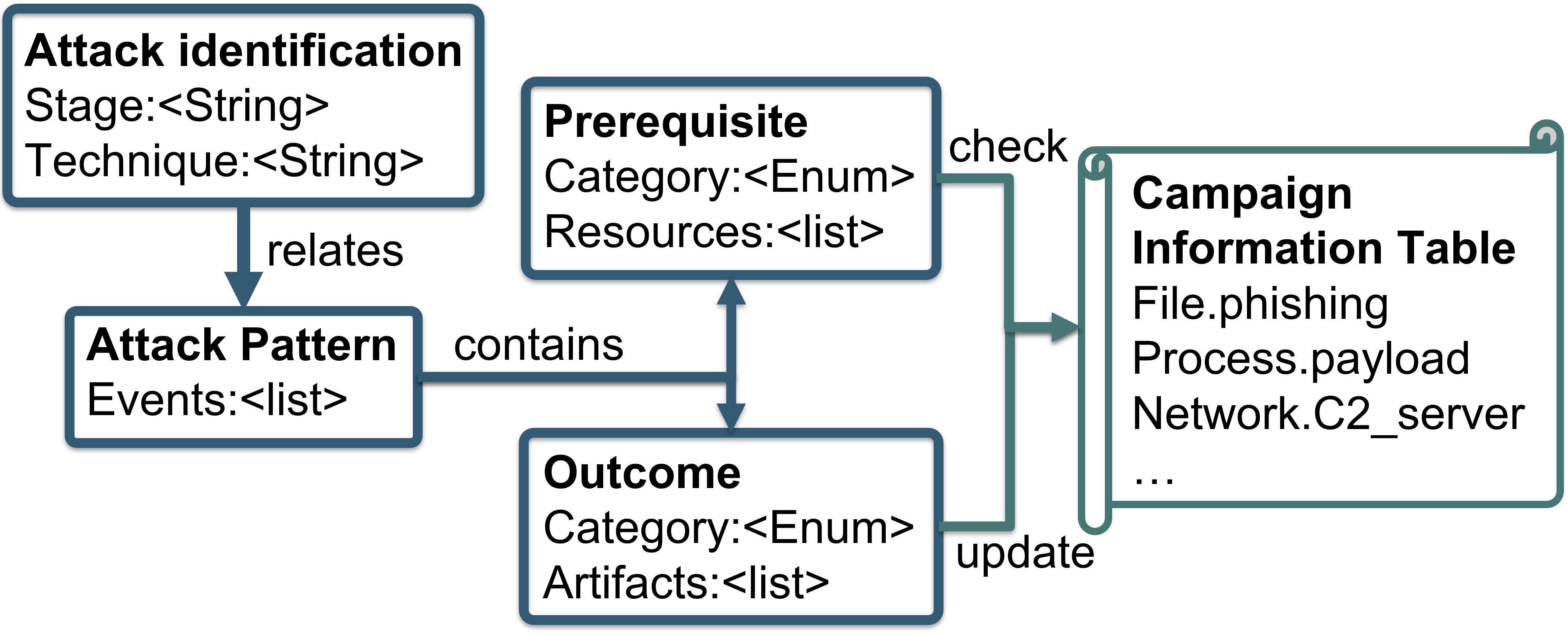}
    \caption{Attack pattern template model}
    \label{fig:template_model}
\end{figure}

\begin{figure}[tb]
    \centering
    \includegraphics[width=0.48\textwidth]{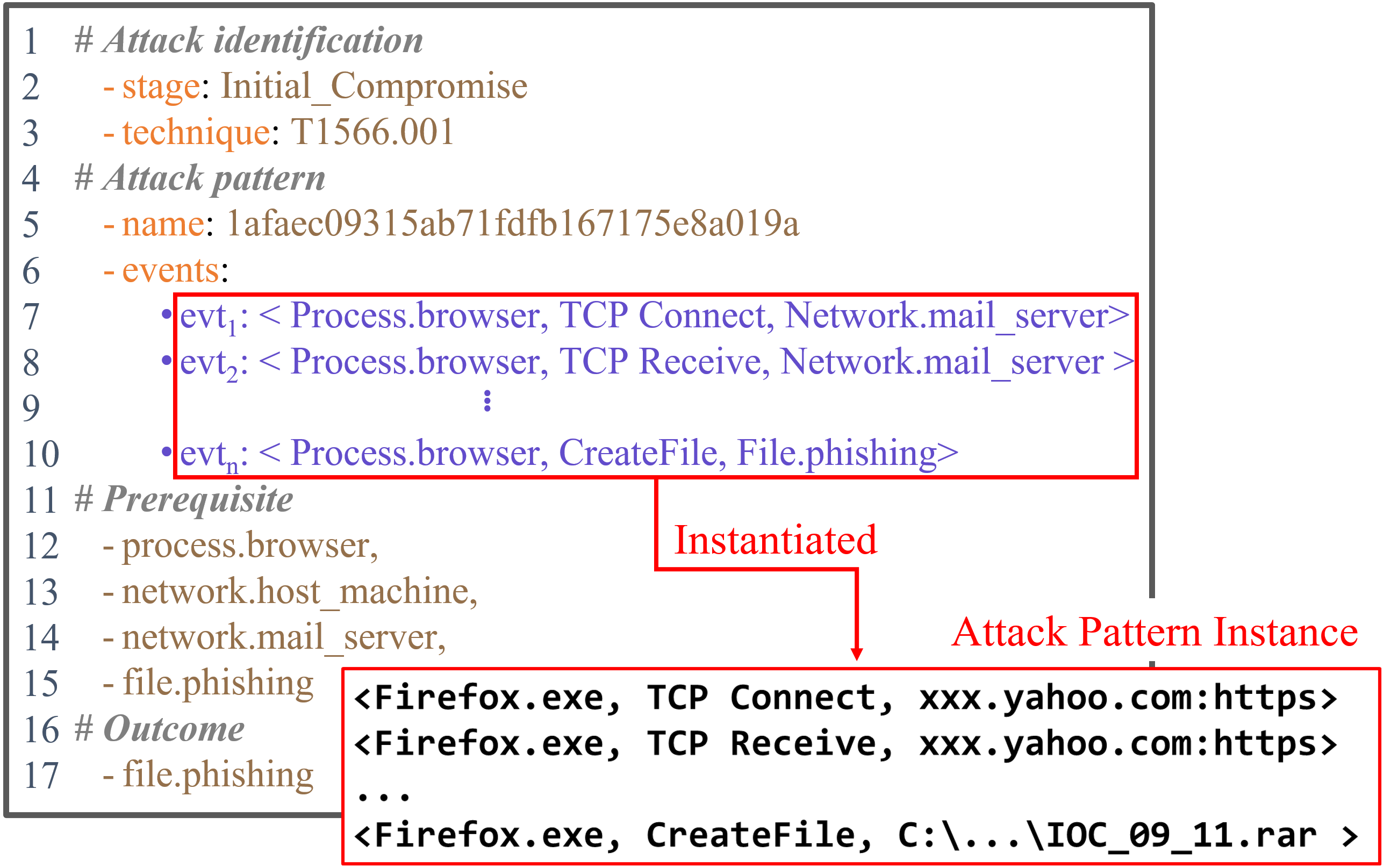}
    \caption{APT28 Attack pattern template example}
    \label{fig:template}
\end{figure}

We abstract each attack pattern into an attack template, which specifies its prerequisites, outcomes, timings, and the corresponding stage and Technique in the attack lifecycle. 
This meta-information serves as a blueprint for generating diverse attack instances while maintaining consistency among the generated instances.
Figure~\ref{fig:template_model} shows the attack template model, which is explained in more detail with an example in Figure~\ref{fig:template}. Algorithm~\ref{alg:generate_template} shows the details of template generation.

\begin{algorithm}
\caption{Generate Attack Pattern Template}
\label{alg:generate_template}
\begin{algorithmic}[1]
\renewcommand{\algorithmicrequire}{ \textbf{Input:}} 
\renewcommand{\algorithmicensure}{ \textbf{Output:}} 
\REQUIRE Attack pattern $A$ and events $E$
\ENSURE Attack pattern template $P$
\STATE Let $TAB$ that stores known system entities and their descriptors and categories, as defined in Table~1.
\STATE Initialize $P$ with the stage and Technique, as described in Subsection "Attack Identification".
\FOR{each event $e$ in $E$}
    \FOR{each system entity $s$ in $E$}
        \IF{$s$ already used}
            \STATE Replace $s$ with its descriptor $d$ from $TAB$.
        \ELSE
            \STATE Find its descriptor $d$ and replace $s$.
            \STATE Set the descriptor as a prerequisite.
        \ENDIF
        \STATE {Set $d$ as an outcome if it still exists.}
        \STATE Insert $e$ into $P$.
    \ENDFOR
\ENDFOR        
\RETURN $P$
\end{algorithmic}
\end{algorithm}

\noindent \textbf{Attack Identification}. 
The entity \textit{Attack Identification} includes the attack \textit{stage} and \textit{Technique}, which together define the identity of the attack pattern template, as shown in Figure~\ref{fig:template_model}. For example, lines 1-3 of Figure~\ref{fig:template} illustrate an attack pattern template associated with Technique \textit{T1566.001 Spearphishing Attachment} and the stage \textit{Initial Compromise} in the adversary lifecycle.

\noindent \textbf{Attack Pattern}. 
The entity \textit{Attack Pattern} is associated with an attack identity and consists of a sequence of events generated by a specific ability, categories and descriptors of artifacts presented in the event. In this context, the category provides a broad classification, while the descriptor offers a more specific identification. For example, if an event contains the artifact \textit{Firefox}, it falls into the category \textit{process} with the descriptor \textit{browser}. To support the generalization of attack pattern templates, Table~\ref{tab:hypernym} outlines a collection of categories and descriptors used in SAGA, which can be considered as forming a hypernym relationship. The table outlines six categories, File, Network, Cmdline, Process, Registry, and System, each subdivided into specific descriptors.
Refer to lines 3-6 of Algorithm~\ref{alg:generate_template}.

Subsequently, additional attack instances can be generated by instantiating these attack pattern templates during the generation of synthetic audit logs.
Using the same running example, three system entities (artifacts) are involved in phishing: \textit{Firefox} accessing \textit{Microsoft Exchange} and downloading \textit{IOC\_09\_11.rar} through the browser. These entities can be substituted with hypernyms (as shown in line 12 of Algorithm~\ref{alg:generate_template} and lines 7-10 of Figure~\ref{fig:template}), such as replacing \textit{Firefox} with \textit{Process.browser} and \textit{IOC\_09\_11.rar} with \textit{File.phishing}.

\noindent \textbf{Prerequisite}. 
The entity \textit{Prerequisite} specifies conditions related to the system resources required for the execution of an attack pattern(lines 8-9 of Algorithm~\ref{alg:generate_template}), such as \textit{process.browser} and \textit{file.phishing} (as shown in lines 12-15 of Figure~\ref{fig:template}) of the attack pattern template T1566.001 of APT28. To generate a specific malicious audit log (i.e., an attack pattern), a campaign information table records previously executed attacks and the system entities utilized, as well as the system entities provided by the external environment, to facilitate \textit{Prerequisite} matching. An attack pattern is generated only when all \textit{Prerequisite} are satisfied; more details are given in Section~\ref{sec: template selection}. 
    
\noindent \textbf{Outcome}. 
The entity \textit{Outcome} lists the artifacts produced after executing this attack
(line 11 of Algorithm~\ref{alg:generate_template}), such as a phishing file (as indicated on line 17 of Figure~\ref{fig:template}), collected sensitive information, and changes in system state. In a complete APT attack chain, \textit{Outcome}  may serve as prerequisites for subsequent attack patterns. Therefore, defining the \textit{Outcome} within the template is crucial, as it not only impacts the immediate effects of the attack, but also has long-term implications for future attack strategies and defensive measures.

\begin{table}[tb]
\caption{Categories and descriptors used in this study.}
\label{tab:hypernym}
\centering
\setlength{\tabcolsep}{1mm}{
\begin{tabular}{|l|l|l|l|}
\hline
Category                 & Descriptor          & Category                  & Descriptor    \\ \hline
\multirow{8}{*}{File}    & ADS(Alter. data streams)                 & \multirow{4}{*}{Process}  & Browser*      \\ \cline{2-2} \cline{4-4} 
                         & Exfiltration Folder &                           & Explorer*     \\ \cline{2-2} \cline{4-4} 
                         & Payload             &                           & Payload       \\ \cline{2-2} \cline{4-4} 
                         & Payload Copy        &                           & Phishing      \\ \cline{2-4} 
                         & Phishing            & \multirow{2}{*}{Registry} & Key           \\ \cline{2-2} \cline{4-4} 
                         & Recon               &                           & Subkey        \\ \cline{2-4} 
                         & Script              & \multirow{11}{*}{System}  & Firewall Port*\\ \cline{2-2} \cline{4-4} 
                         & Shortcut            &                           & Firewall Rule \\ \cline{1-2} \cline{4-4} 
\multirow{6}{*}{Network} & C2                  &                           & Host*         \\ \cline{2-2} \cline{4-4} 
                         & Host IP*            &                           & Password*     \\ \cline{2-2} \cline{4-4} 
                         & Host Machine*       &                           & Product       \\ \cline{2-2} \cline{4-4} 
                         & Mail Server         &                           & Proxy Port*   \\ \cline{2-2} \cline{4-4} 
                         & Payload URL         &                           & Service       \\ \cline{2-2} \cline{4-4} 
                         & Script URL          &                           & Task          \\ \cline{1-2} \cline{4-4} 
\multirow{3}{*}{Cmdline} & Command             &                           & Time*         \\ \cline{2-2} \cline{4-4} 
                         & Content             &                           & User*         \\ \cline{2-2} \cline{4-4} 
                         & Message             &                           & Userdomain    \\ \hline
\end{tabular}}
    \begin{tablenotes}
        \item * indicates the instances of descriptor are generated by using Faker~\cite{faker}.
    \end{tablenotes}
\end{table}

\subsection{APT Lifecycle and Technique Selection}\label{sec: Synthesized_event}

SAGA supports both random and user-specified APT campaign generation, and in this subsection, the random APT campaign generation is introduced. The generation of an APT campaign is conditioned on the adversary lifecycle, involving lifecycle stage transitions, attack template selections, and attack pattern instantiation. 
The transition process of the lifecycle stage determines the subsequent stage according to the current status.
An example of an adversary Mandiant lifecycle is shown in Figure~\ref{fig:Running_example} \circled{a}. 
The attack template selection process chooses the attack templates for each stage, while the artifact substitution process instantiates the events for an attack pattern.

\subsubsection{Lifecycle Stage Transitions}
As described in Section~\ref{sec: lifecycle}, threat actors follow a systematic, strategic approach, executing actions sequentially. 
An APT campaign begins with two initial stages: \textit{initial compromise} (IC) and \textit{establish foothold} (EF).  This is followed by an incubation period consisting of four stages: \textit{escalate privileges} (EP), \textit{internal reconnaissance} (IR), \textit{move laterally} (ML), and \textit{maintain presence} (MP). The incubation period may occur once or multiple times, and within each incubation period, individual stages may be omitted or repeated. Finally, the last stage, \textit{complete mission} (CM), is executed once or not at all, marking the successful completion of the attack. 
The Context-Free Grammar representation of a campaign generation is presented in Table~\ref{tbl:CFG}. The lifecycle is defined in line 6.
As an example, in Figure~\ref{fig:Running_example} \circled{b}, the stages of APT28 include Initial Compromise, Establish Foothold, Internal Reconnaissance, and Complete Mission.

\newcounter{rownum}
\setcounter{rownum}{0}
\newcommand{\rownumber}{\stepcounter{rownum}\therownum. }

\begin{table}[tb!]
\centering
\caption{Context Free Grammar for Campaign Generation}\label{tbl:CFG}
\renewcommand{\arraystretch}{1.2}
\setlength{\tabcolsep}{1mm}{
\begin{tabular}{|p{0.3cm}l|}
\hline

\rownumber &  $SA$ :- $C_ {m} (SA,MA)$ $\vert$ $C_ {b} (SA,BA)$ $\vert$ $MA$ $\vert$ $BA$ $\vert$ $\emptyset$  \\

\rownumber &  $BA$ :- $EVT$ \\

\rownumber &  $EVT$  :-  $evt+EVT$ $\vert$ $evt$ $\vert$ $\emptyset$ \\

\rownumber & \\ 
\rownumber & /* O+ : zero time or more, 1+ : one time or more,\\
           & 0,1 : zero or one time */  \\
\rownumber &  $LC$ :- $IC+EF+(EP^{0+}+IR^{0+}+ML^{0+}+$ \\
           & $MP^{0+})^{1+}+CM^{0,1}$ \\
\rownumber & \\

\rownumber &  $MA$ :- $BE+In(TQ)+MA$ $\vert$ $BE+In(TQ)$ \\
\rownumber &  $BE$ :- $EVT$ \\

\rownumber &  $In(TQ)$ :- $In(TQ)$ if TQ.prerequisites are satisfied $\vert$ $\emptyset$ \\

\hline
\end{tabular}}
    \begin{tablenotes}
        \item \textbf{Notation:} 
        $SA$ = synthetic audit log,  
        $MA$ = malicious audit log,  
        $BA$ = benign audit log, 
        $LC$ = APT lifecycle,     
        $IC$ = initial compromise,
        $EF$ = establish foothold,
        $EP$ = escalate privileges, 
        $IR$ = internal reconnaissance, 
        $ML$ = move laterally,
        $MP$ = maintain presence,
        $CM$ = complete mission,        
        $BE$ = a sequence of benign events,
        $EVT$ = event sequence,  
        $TQ$ = attack template,  
        $In(TQ)$ = attack pattern instance         
    \end{tablenotes}
\end{table}

\subsubsection{Attack Template Selection}\label{sec: template selection}

For each stage of an APT campaign, the number of attacks can be either randomly determined or specified by the user. An attack template (TQ) may also be randomly selected based on the current stage and prerequisites, or as specified by the user. 
Note that there may be benign events (BE) preceding the attack pattern templates (TQ) that the operation is described in line 8 of Table~\ref{tbl:CFG}. 
In line 8, malicious audit logs (MA) consist of either a sequence of benign events (BE), the selected attack pattern instance ($In(TQ)$), malicious audit logs (MA), 
or sequences of benign events (BE) and the selected attack pattern instance ($In(TQ)$).
Next, the abstracted arguments of the template are replaced with artifacts to generate diverse audit events, as the instantiation of an attack pattern described in line 10 ($In(TQ)$) and elaborated in Subsection~\ref{sec: substitution}. 
The construction of the malicious audit log continues until all stages have been completed.

A campaign information table, as depicted in Figure~\ref{fig:template_model}, is created to maintain information about the executed attack templates, including their associated lifecycle stage, as well as the system entities used and their statuses. 
Given the current stage, the prerequisites of a template are examined against the campaign information table to determine whether they are satisfied. For example, if a threat actor uses a phishing email template during the initial compromise stage, the campaign information table will record the details of the phishing email template, including the entity \textit{File.phishing}. In the second stage (EF), the templates for this stage will refer to the campaign information table to check whether their prerequisites are met. Only if the prerequisites are met can they be considered potential templates for this APT campaign.

\renewcommand{\algorithmicrequire}{ \textbf{Input:}} 
\renewcommand{\algorithmicensure}{ \textbf{Output:}} 

\begin{algorithm}[tb!]
\caption{Synthetic Audit Log Generation}
\label{alg:campaign_gen}
\begin{algorithmic}[1] 
\REQUIRE ~~ 
    Lifecyle $\{LC\}$,
    Attack pattern instances $\{ln(TQ)\}$,
    Benign audit log $\{BA\}$
    \ENSURE ~~ 
    Synthetic audit events $SA$

\WHILE {while $LC_i$ in {$LC$} exists}    
\FOR{each $stage$ in $LC_i$}
  \STATE {// Include an attack pattern instance $ln(TQ)$ in the $stage$. Line 8 of Table~\ref{tbl:CFG}}.
    \STATE $MA$ = $BE+In(TQ)+MA$ $\vert$ $BE+In(TQ)$
  \STATE {where $In(TQ)$ is an instantiated attack template, as Line 10 of Table~\ref{tbl:CFG}. $BE$ is a short list of benign events.}

\ENDFOR
\ENDWHILE {~~//Obtain all malicious audit logs $\{ MA \}$}
\STATE
\STATE Initialize $SA$ with given background events.
\FOR{each $BA_i$ in $\{BA\}$}
\STATE {// Perform $C_{b}(SA, BA_i )$, as Line 1 of Table~\ref{tbl:CFG}}.
\FOR{each $evt$ in $\{ BA_i \}$}
\STATE $evt.start = {BA_i}.start + evt.relative$
\ENDFOR {~// Insert event $evt$ to synthetic audit log.}
\STATE $SA$ = $SA \cup BA_i$
\ENDFOR {~// Insert benign audit log $BE_i$.}

\FOR{each $MA_i$ in $\{MA\}$}
\STATE {// Perform $C_{m}(SA,MA)$, as Line 1 of Table~\ref{tbl:CFG}}.
\FOR{each $In(TQ_j)$ in $\{ MA_i \}$}
\FOR{each $evt$ in $\{In(TQ_j)\}$}
\STATE $evt.start = In(TQ_{j-1}).end + In(TQ_{j}).lapse + evt.relative$
\ENDFOR {~// Insert event $evt$ to synthetic audit log.}
\STATE $SA$ = $SA \cup In(TQ_j)$
\ENDFOR {~// Insert technique $In(TQ)$ to synthetic log.}
\ENDFOR {~// Insert malicious audit log $MA$ to synthetic audit log.}  

\end{algorithmic}
\end{algorithm}

\subsubsection{Attack Pattern Instantiation}\label{sec: substitution}

In Subsection \ref{sec: template}, we discuss the creation of attack templates through the generalization of system entities. 
While instantiating an attack pattern template, these generalized system entities are replaced by appropriate and meaningful artifacts (or a hyponym) to form a validated attack pattern. This approach enables the generation of a wide range of attack instances.

There are two methods for collecting the hyponyms of generalized system entities. The first method involves using artifacts collected from real-world data, such as malware names, attack-related file names, and registry paths. To gather these artifacts, we leveraged VX-Underground~\cite{vxheaven} and VirusTotal~\cite{virustotal}. 
VX-Underground provides approximately 10,000 malware hash values associated with APT campaigns collected over the past five years. Each hash was subsequently queried on VirusTotal to extract relevant artifacts from the behavioral analysis reports. In total, 157,806 artifacts were obtained and mapped to the generalized system entities defined in the attack templates for subsequent instantiation.
For example, by querying VirusTotal using a sample hash value 78a3e4702d9fc13b2ef917211cd65b44, we can extract a file \textit{\%APPDATA\%\textbackslash $...$ \textbackslash Startup\textbackslash twainResolver.lnk}. The dropped file, \textit{twainResolver.lnk}, is classified under the system entity type \textit{file}, as indicated in the `file' section of the VirusTotal report.
It is then recognized as `file.short' based on regular expression matching, and a shortcut file in the Windows startup folder to implement an attack behavior such as \textit{T1547.009 Shortcut Modification}.

The second source of artifacts, such as usernames and process IDs, which are indicated with $*$ Table~\ref{tab:hypernym}, is generated using the Faker library in Python~\cite{faker}. Faker efficiently generates realistic-looking random data that mimics real-world information. These random artifacts are used to populate attack instances within the templates.
Once attack patterns are generated for all stages as a chronological sequence of labeled events, the prerequisites for every attack pattern must be satisfied to validate the attack sequence, which is then considered an APT campaign.

\subsection{Synthetic Audit Log Generation}\label{sec: event_insertion}
A synthetic audit log comprises several benign and APT campaign audit logs, with their composition methods detailed in Algorithm~\ref{alg:campaign_gen}.
A benign audit log is recorded from normal software activities, such as web browsing, Windows Office usage, file downloads, etc., and is represented as a sequence of events, as defined in lines 2 and 3 of Table~\ref{tbl:CFG}. Each benign audit log includes a start time (BA.start) and log length (BA.length), with each event having a relative time (evt.relative) to the log start time. The notation $C_{b}(SA,BA)$ in the first line of Table~\ref{tbl:CFG} denotes the composition of a benign audit log $BA$ into a synthetic audit log $SA$. The composition involves setting the starting time of the first event to the specified time of the benign audit log, and adjusting the start times of all other events by adding the log start time to their relative times. The procedure is described in lines 9 to 14 of Algorithm~\ref{alg:campaign_gen}.

For the composition of an APT campaign audit log into a synthetic audit log, denoted as $C_{m}(SA,MA)$ in the first line of Table~\ref{tbl:CFG}, it merges each pair of benign events (BE) and attack technique events ($ln(TQ)$) sequentially until all pairs are combined. A $BE$ is usually a short sequence of benign events that occurs before the particular Technique ($ln(TQ)$). The composition of $BE$ follows the same method as that for $BA$. The composition of $TQ$, which refers to the audit log events representing a particular Technique, provides flexibility in both the start time and duration. This means that a Technique audit log can begin at any time and span any length. The start time of a Technique audit log is set to the end time of the preceding Technique plus the lapse time $ln(TQ).lapse$, which can be specified by the user or determined randomly by SAGA. The procedure is described in lines 15 to 22 of Algorithm~\ref{alg:campaign_gen}. 

\begin{figure}[t!]
    \centering
    \includegraphics[width=0.45\textwidth]{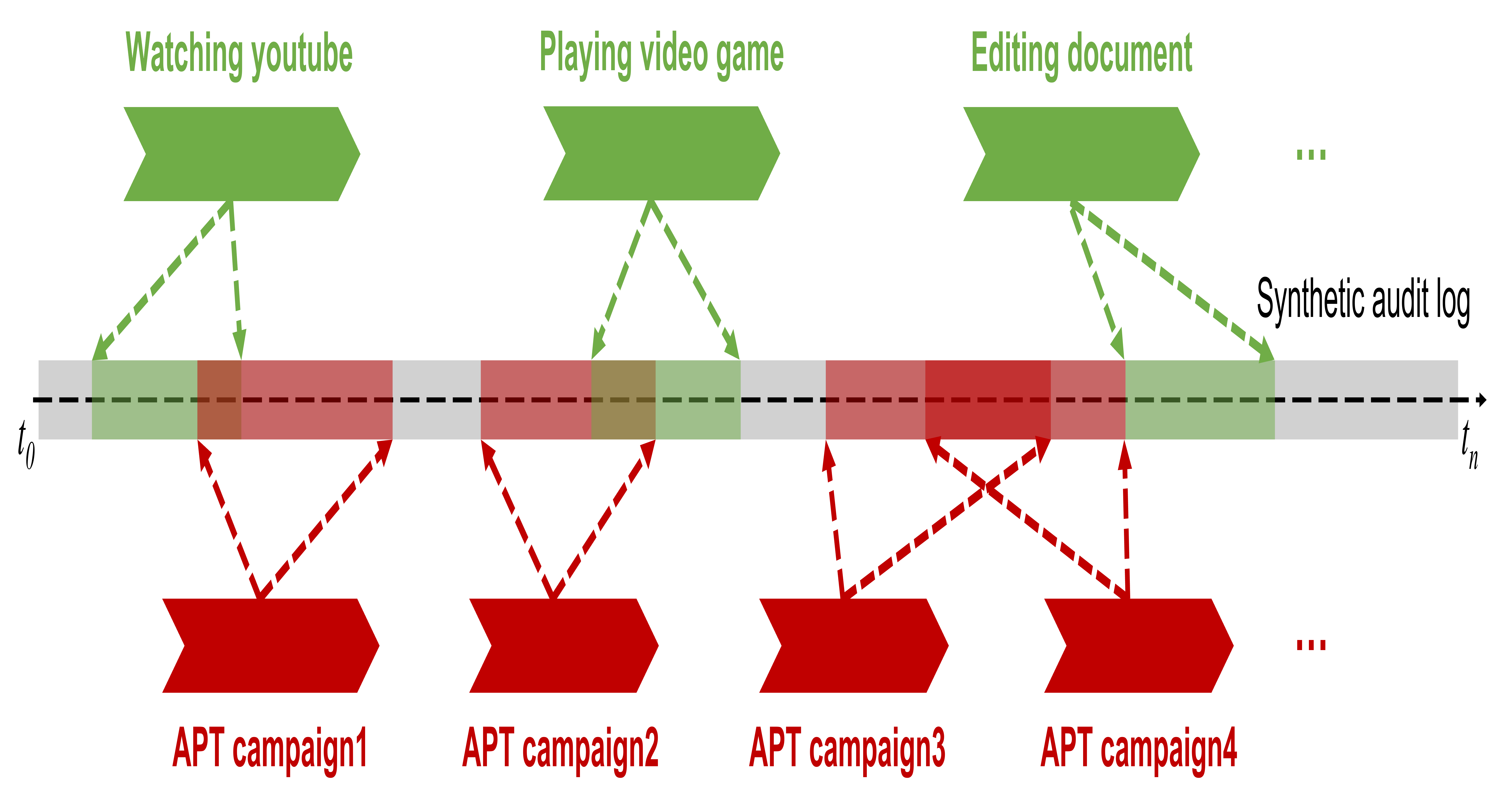}
    \caption{Composition of Synthetic audit log}
    \label{fig:Synthetic_log}
\end{figure}

Figure~\ref{fig:Synthetic_log} depicts a composition of 3 benign audit logs and 4 APT audit logs.
Note that the logs may overlap and the APT audit logs can be extended for any duration.
SAGA explicitly leverages process identifiers, parent-child process relationships, and event timestamps to preserve the underlying process structure, temporal ordering, and causal dependencies that are critical for contextual consistency in audit logs.
To avoid unrealistic combinations of events, SAGA maintains semantic coherence at both the process and artifact levels. 
For example, when two events involve shared entities, such as a file, registry key, or named pipe, SAGA preserves these dependencies to ensure that the resulting logs reflect plausible and continuous behavior. 
As illustrated in Subsection II-C, a user downloads a malicious attachment IOC\_09\_11.rar via a browser (\textit{T1566.001: Spearphishing Attachment}), and subsequently opens it using Microsoft Word (\textit{T1204.002: Malicious File Execution}). 
Although these steps span separate processes, their interaction through a shared artifact forms a coherent and realistic attack sequence.
In this way, SAGA ensures that injected attack behaviors are not only chronologically plausible but also contextually consistent with normal system operations, mitigating the risk of generating disjointed or unrealistic event flows.

There are still tasks that need to be addressed, such as establishing process relationships. For example, in an audit log, the ProcMon operation \textit{ProcessCreate} (as shown in Figure~\ref{fig:audit_log}) creates a new process, and their parent-child relationship must be preserved. SAGA treats artifacts with the same descriptor within a single attack pattern template as belonging to the same process. Otherwise, operations from different templates are considered independent processes.

\section{Usefulness Study of Synthetic Audit Log}

In this section, we evaluate the usefulness of synthetic audit logs for various detection methods in different APT attack scenarios. 
First, we show the variety of synthetic audit logs that SAGA can produce, from those that mimic real-world APT attacks (denoted as the C Group) to randomly generated logs (denoted as the G group), and finally to composite both groups (denoted as the M group). Next, we assess whether state-of-the-art audit log-based and provenance-based malicious detection methods can effectively use the synthetic audit log as their target datasets. In addition, we discern the performance and operation differences of these methods when applying their original datasets versus various synthetic data logs.
Note that the objective of this study is not to identify the best-performing methods or models, as each approach has its own specific focus, design considerations, and inherent limitations.

This section begins by outlining the settings for the empirical study, including the synthesized datasets, labeling, metrics, and machine hardware and software configurations. 
Our analysis focuses on three sets of empirical studies to demonstrate the usefulness of the synthesized attack datasets:
1) assessing the effectiveness of APT intrusion detection by,
2) evaluating the accuracy of Technique hunting,
and 3) gauging the effectiveness of campaign attribution.
Finally, we will conduct a case study to evaluate the effect of model training using synthetic audit logs to test the model on an unseen dataset.

\subsection{Evaluation Settings}
\noindent \textbf{Datasets.}
The experiments utilize synthetic audit log datasets generated by SAGA (hereafter referred to as the SAGA dataset),
which is publicly available at ~\cite{saga_dataset}. A synthetic audit log, as described in Section~\ref{sec: generativeThreatModel}, includes both benign behaviors and malicious APT attack scenarios. Similarly to previous research~\cite{manzoor2016steampot, alsaheel2021atlas}, the benign audit logs cover a variety of everyday activities, such as watching YouTube videos, downloading and compiling source code from GitHub, writing and sending emails, running Python programs, playing video games, reading news articles, and composing Word documents. This diversity ensures that the synthetic audit logs accurately reflect typical user behavior.
For malicious events, we prepared 169 abilities executed on the MITRE CALDERA~\cite{caldera} red-team emulator to generate labeled malicious audit events. These events were then processed into attack pattern templates for further synthetic audit log generation. Each of the 169 attack pattern templates was instantiated with artifact replacements, as described in Section~\ref{sec: template}, to form the datasets used in the empirical study.
The training dataset for Technique hunting consists of 16,900 attack patterns, each of which is an instantiation of a attack pattern template. This dataset is split into training, validation, and test sets in an 8:1:1 ratio. 

Given the limited availability of audit logs labeled with Techniques, we used the attack pattern templates to create APT campaigns. These scenarios include eight APT attack campaigns (``known'' group) based on cyber threat intelligence reports from real-world APT campaigns, including Higaisa~\cite{Malwarebytes2020Higaisa}, admin338~\cite{Mandiant2015admin@338}, APT28~\cite{APT28}, FIN7~\cite{Mandiant2017FIN7}, CobaltGroup~\cite{ptsecurity2017Cobalt}, Gamaredon~\cite{CERTEE2021Gamaredon}, Patchwork~\cite{Cymmetria2016Patchwork}, and GorgonGroup~\cite{Unit2017Gorgon}. 
Moreover, using SAGA random generation capability, we produced 20 APT campaigns (``random generated group''). From these 28 APT campaigns, we randomly selected 3 to create new composite scenarios involving multiple APT campaigns targeting a single victim host, resulting in 10 composite campaigns (``composite'' group). To demonstrate the versatility and flexibility of SAGA, each of the above APT campaigns was generated over three different time periods: 15 minutes, 1 hour, and 1 day.

\noindent \textbf{Labeling.}
For each generated APT campaign, SAGA labels malicious events with APT stages, Techniques, abilities, manipulated system entities, and other operation-related attributes. In addition, it includes threat actors, when applicable. 
Malicious audit events are labeled using the BIO2 scheme~\cite{sang2003introduction} covering 7 attack stages, 80 Techniques, and 169 abilities, as shown in Table S1 of Supplementary Material A.

\noindent
\textbf{Evaluation Metrics.}
We adopt three evaluation metrics for different tasks: APT intrusion detection, Technique hunting, and APT campaign attribution. The APT intrusion detection task aims to identify any APT intrusions or anomalies within the dataset under examination. Technique hunting seeks to identify Techniques violated based on MITRE ATT\&CK definitions. APT campaign attribution aims to determine the most probable APT campaign present in the dataset. To evaluate performance, we use standard classification metrics: precision (P), recall (R), F1 score (F1), and accuracy (ACC). 

For APT intrusion detection, we frame it as a binary classification problem, determining whether a malicious event exists in the audit log. Since different target systems or methods apply their own tests to detect APT intrusion on an event, a graph, or a time window interval, we adapt to their respective evaluation designs to assess performance. Technique hunting is treated as a multi-class classification task, where each classifier determines whether a specific Technique has been employed in one or more events. For APT campaign attribution, we analyze the Techniques identified from the Technique hunting task to determine the most likely APT campaign for the audit log. We rank the ground truth against the predictions and employ a Top-K measurement to evaluate the performance of the APT campaign attribution.

\noindent
\textbf{Established Baselines}
For APT intrusion detection, we demonstrate the usefulness of generated attack datasets using three types of systems: event-based (AirTag~\cite{ding2023airtag}), graph-based (Unicorn~\cite{han2020unicorn}), and time window-based (KAIROS~\cite{cheng2023kairos}) intrusion detection systems.
The reasons for choosing the three systems are not only that they are established baselines, but also that their codes are publicly available and the systems can be installed and executed without issues. 

For Technique hunting, we used the Sigma rules~\cite{sigma}, an open signature-based rulebase contributed by numerous security practitioners, and SFM~\cite{huang2024cascadeapproachaptcampaign}, a deep learning approach that provides Technique hunting functionality. We selected Sigma as a baseline because it is a widely adopted, community-driven detection framework, actively maintained by practitioners worldwide. As noted by ~\cite{virkud2024does}, Sigma has also been shown to align closely with several commercial rulesets, making it a meaningful and representative point of comparison.
A major portion of these rules align with the MITRE ATT\&CK framework with 70 rules mapped to the Techniques identified in this study.
SFM integrates three key technologies: anomaly detection to mitigate the impact of data imbalance, a multi-class classifier for Technique hunting, and subgraph matching for APT campaign attribution.
SFM was selected because it was one of the few systems specifically designed to address both the Technique hunting task and APT campaign attribution, making it particularly suited for the evaluation of complex attack scenarios.
All models and methods were installed with configurations as close as possible to those described in their original publications. Code modifications were only made to ensure compatibility with the structure of the synthetic audit logs.

\begin{table}[tb!]
\caption{Computation overhead of Audit Log Generation.}
\label{tab:overhead}
\centering
\setlength{\tabcolsep}{1mm}{
\begin{tabular}{|l|l|lll|}
\hline
\multirow{2}{*}{}                      & \multicolumn{1}{c|}{\multirow{2}{*}{\begin{tabular}[c]{@{}c@{}}Synthesizing \\APT Campaign \end{tabular}}} & \multicolumn{3}{c|}{Audit Log Composition}                                          \\ \cline{3-5} 
                                       & \multicolumn{1}{c|}{}                                                                                      & \multicolumn{1}{c|}{15 min}     & \multicolumn{1}{c|}{1 hour}    & \multicolumn{1}{c|}{1 day} \\ \hline
\multicolumn{1}{|l|}{Memory (MB)} & \multicolumn{1}{r|}{46.95}                                                                                 & \multicolumn{1}{r|}{2644.05}   & \multicolumn{1}{r|}{4407.96}  & \multicolumn{1}{r|}{41078.54}                         \\ \hline
\multicolumn{1}{|l|}{Time (Sec)}   & \multicolumn{1}{r|}{54.87}                                                                                 & \multicolumn{1}{r|}{47.07}     & \multicolumn{1}{r|}{76.61}    & \multicolumn{1}{r|}{322.84}                          \\ \hline
\end{tabular}}
\end{table}

\noindent
\textbf{Computation Configurations.}
For empirical studies, we used a Windows 10 operating system with 8GB of memory with an Intel i9-10900F CPU as our victim system. For defender model development, we employed AMD EPYC 7282 16-Core Processor CPU running Ubuntu 20.04 system with 1TB of memory and two NVIDIA A100 80GB GPUs. Neural network training was performed using PyTorch 1.13.1 and Python 3.10.
The computational overhead of generating malicious audit events and audit log composition was measured and listed in Table~\ref{tab:overhead}.

\begin{table*}[!htb]
\caption{APT attack scenarios based on cyber threat intelligence reports.}
\label{tab:known_campaigns}
\centering
\begin{tabular}{|l|l|l|l|}
\hline
ID  & Simulated Campaign                        & Attack Stage          & Techniques                                   \\ \hline
C1  & Higaisa~\cite{Malwarebytes2020Higaisa}    & \{1,2,6,4,4,6,6\}     & PA, MFE, RK3, SID, SNCD3, MTOS1, ST2         \\ \hline
C2  & admin338~\cite{Mandiant2015admin@338}     & \{1,2,4,4,4,4,4,4\}   & PA, MFE, LA1, FDD1, LG1, SNCD6, SNCoD1, SSD3 \\ \hline
C3  & APT28~\cite{APT28}                        & \{1,2,2,4,4,7\}       & PA, WP, MFE, SID, DLS, EWS                  \\ \hline
C4  & FIN7~\cite{Mandiant2017FIN7}              & \{1,2,6,6\}           & PA, ITT3, RK4, ST1                           \\ \hline
C5  & CobaltGroup~\cite{ptsecurity2017Cobalt}   & \{1,2,4\}             & PA, RAS1, NSD                                \\ \hline
C6  & Gamaredon~\cite{CERTEE2021Gamaredon}      & \{1,2,2,6,6,4,4,6,7\} & PA, WP, MFE, MR3, RK2, WMI5, SID, ST2, DF2   \\ \hline
C7  & Patchwork~\cite{Cymmetria2016Patchwork}   & \{1,2,3,4,4,4,6,5\}   & PA, PS2, BUAC, DLS, UD2, SD1, RK3, RDP       \\ \hline
C8  & GorgonGroup~\cite{Unit2017Gorgon}         & \{1,2,6,6,6,6,6\}     & PA, PS1, PEI, RK3, SM1, DMT, HW              \\ \hline
\end{tabular}
    \begin{tablenotes}
        \item Lifecyle is indexed by number. Techniques include PA = phishing Attachment, MFE = Malicious File Execution, RK = Registry Run Keys, SID = System Information Discovery, SNCD = System Network Configuration Discovery, MTOS = Masquerade Task or Service, ST = Scheduled Task, LA = Local Account, FDD = File and Directory Discovery, LG = Local Groups, SNCoD = System Network Connections Discovery, SSD = System Service Discovery, WP = Web Protocols, EWS = Exfiltration Over Web Service, ITT = Ingress Tool Transfer, RAS = Remote Access Software, NSD = Network Service Discovery, MR = Modify Registry, WMI = Windows Management Instrumentation, DF = Defacement, PS = PowerShell, BUAC = Bypass User Account Control, DLS = Data from Local System, UD = System Owner/User Discovery, SD = Security Software Discovery, RDP = Remote Desktop Protocol, PEI = Portable Executable Injection, SM = Shortcut Modification, DMT = Disable or Modify Tools, HW = Hidden Window. The subsequent number of a technique represents a distinct ability used to implement that technique.
    \end{tablenotes}
\end{table*}

\begin{table*}[!htb]
\caption{Statistics for the three time durations in the eight APT attacks.}
\label{tab:known_campaigns_stat}
\centering
\begin{tabular}{|c|r|rrr|rrr|rrr|}
\hline
\multirow{2}{*}{ID} & \multicolumn{1}{c|}{\multirow{2}{*}{\begin{tabular}[c]{@{}c@{}}\# Malicious \\ event\end{tabular}}} & \multicolumn{3}{c|}{15 min}                                                                     & \multicolumn{3}{c|}{1 hour}                                                                     & \multicolumn{3}{c|}{1 day}                                                                      \\ \cline{3-11} 
                           & \multicolumn{1}{c|}{}                                                                               & \multicolumn{1}{c|}{size(MB)} & \multicolumn{1}{c|}{\# entity} & \multicolumn{1}{c|}{\# event} & \multicolumn{1}{c|}{size(MB)} & \multicolumn{1}{c|}{\# entity} & \multicolumn{1}{c|}{\# event} & \multicolumn{1}{c|}{size(MB)} & \multicolumn{1}{c|}{\# entity} & \multicolumn{1}{c|}{\# event} \\ \hline
C1                         & 30                                                                                                  & \multicolumn{1}{r|}{303.27}   & \multicolumn{1}{r|}{31,193}    & 609,394                       & \multicolumn{1}{r|}{1,344.68} & \multicolumn{1}{r|}{255,180}   & 2,400,089                     & \multicolumn{1}{r|}{7,423.86} & \multicolumn{1}{r|}{696,844}   & 14,647,429                    \\ \hline
C2                         & 53                                                                                                  & \multicolumn{1}{r|}{472.79}   & \multicolumn{1}{r|}{34,931}    & 956,654                       & \multicolumn{1}{r|}{1,304.79} & \multicolumn{1}{r|}{138,851}   & 2,609,138                     & \multicolumn{1}{r|}{7,394.81} & \multicolumn{1}{r|}{747,445}   & 14,548,395                    \\ \hline
C3                         & 14,137                                                                                              & \multicolumn{1}{r|}{684.40}   & \multicolumn{1}{r|}{182,004}   & 1,209,895                     & \multicolumn{1}{r|}{179.00}   & \multicolumn{1}{r|}{33,546}    & 360,313                       & \multicolumn{1}{r|}{7,447.71} & \multicolumn{1}{r|}{709,641}   & 14,682,885                    \\ \hline
C4                         & 22                                                                                                  & \multicolumn{1}{r|}{1,161.21} & \multicolumn{1}{r|}{276,956}   & 2,072,382                     & \multicolumn{1}{r|}{137.75}   & \multicolumn{1}{r|}{14,912}    & 290,962                       & \multicolumn{1}{r|}{7,414.71} & \multicolumn{1}{r|}{696,483}   & 14,631,215                    \\ \hline
C5                         & 1,133                                                                                               & \multicolumn{1}{r|}{483.48}   & \multicolumn{1}{r|}{36,466}    & 969,542                       & \multicolumn{1}{r|}{730.36}   & \multicolumn{1}{r|}{73,220}    & 1,475,802                     & \multicolumn{1}{r|}{6,622.55} & \multicolumn{1}{r|}{667,110}   & 12,914,561                    \\ \hline
C6                         & 59                                                                                                  & \multicolumn{1}{r|}{225.76}   & \multicolumn{1}{r|}{31,494}    & 453,915                       & \multicolumn{1}{r|}{199.48}   & \multicolumn{1}{r|}{22,316}    & 399,084                       & \multicolumn{1}{r|}{8,002.90} & \multicolumn{1}{r|}{741,686}   & 15,615,065                    \\ \hline
C7                         & 14,124                                                                                              & \multicolumn{1}{r|}{83.51}    & \multicolumn{1}{r|}{29,655}    & 163,487                       & \multicolumn{1}{r|}{800.95}   & \multicolumn{1}{r|}{131,125}   & 1,623,759                     & \multicolumn{1}{r|}{6,415.17} & \multicolumn{1}{r|}{669,041}   & 12,558,106                    \\ \hline
C8                         & 50                                                                                                  & \multicolumn{1}{r|}{494.16}   & \multicolumn{1}{r|}{53,386}    & 850,419                       & \multicolumn{1}{r|}{153.08}   & \multicolumn{1}{r|}{18,312}    & 316,953                       & \multicolumn{1}{r|}{7,375.18} & \multicolumn{1}{r|}{747,088}   & 14,509,490                    \\ \hline
\multicolumn{2}{|r|}{Mean}                                                                                                                             & \multicolumn{1}{r|}{488.57}   & \multicolumn{1}{r|}{84,511}    & 910,711                       & \multicolumn{1}{r|}{606.26}   & \multicolumn{1}{r|}{85,933}    & 1,184,513                     & \multicolumn{1}{r|}{7,262.11} & \multicolumn{1}{r|}{709,417}   & 14,263,393                    \\ \hline
\end{tabular}
\end{table*}

\subsection{APT dataset Generation}
In this section, we provide a more detailed discussion on the considerations for generating the APT dataset, focusing primarily on generation guidance, attack scenarios, and duration.

\noindent \textbf{Generation Guidance.} 
SAGA offers flexibility in generating APT campaigns, based on user specifications, randomly, or through a combination of both. For example, SAGA generates eight known APT campaigns (denoted from C1 to C8) following user specifications, with their lifecycles and Techniques presented in Table~\ref{tab:known_campaigns}.
To further illustrate SAGA's versatility, we also used its random generation capabilities to create 20 additional campaigns (denoted from G1 to G20), showcasing its capacity to create varied attack scenarios. The lifecycles and Techniques of these randomly generated campaigns are presented in Table S2 of Supplementary Material A, while the detailed statistics, including sizes, number of entities, and number of events, are provided in Table S3 also in Supplementary Material A.
This flexibility highlights the scalability and adaptability of our approach in simulating a wide range of cyberattack scenarios, enabling the creation of customized and diverse testing environments to thoroughly evaluate detection methods.

\noindent \textbf{Attack Scenario.}
We introduce two variations in attack scenarios: one involving a single campaign targeting the system and another featuring three simultaneous campaigns. This variation in attack composition allows for a comprehensive assessment of detection methods in more complex, multi-campaign situations. For instance, the M1 attack scenario consists of three separate attacks, Higaisa, G1, and admin338, each employing distinct Techniques and targeting the same compromised host. Additional details on the ten generated composite campaigns can be found in Table S4 in Supplementary Material A, due to the constraints of the manuscript space.

\noindent \textbf{Attack Duration.} 
To account for the variability in the duration of the attack, we generated synthetic audit logs with durations of 15 minutes, 1 hour, and 1 day. Although SAGA can create logs of any length, we limited the maximum duration to 1 day in this study, considering the computational overhead of the intrusion detection systems discussed in Section~\ref{sec: intrusion system}
Table~\ref{tab:known_campaigns_stat}, Table S3 and Table S4 in Supplementary Material A, presents statistical descriptions of logs for each duration. 
As shown in Table~\ref{tab:known_campaigns_stat}, the average number of events increases with log length: 910,711 for 15-minute audit logs, 1,184,512.5 for 1-hour audit logs, and 14,263,393.25 for 1-day audit logs.
This variability reflects the evolving nature of APT attacks, which are not constrained by fixed timeframes, highlighting SAGA ability to generate audit logs of various durations. It also allows for a more thorough examination of how different detection methods perform over varying time intervals.

\begin{table*}[!htb]
\caption{Average performance of the intrusion detection systems}
\label{tab:6_3_average_intrusion_result}
\centering
\setlength{\tabcolsep}{1mm}{
\begin{tabular}{|ll|rrr|rrr|rrr|rrr|r|}
\hline
\multicolumn{2}{|c|}{\multirow{2}{*}{Campaign}}       & \multicolumn{3}{c|}{Precision}                                                              & \multicolumn{3}{c|}{Recall}                                                                 & \multicolumn{3}{c|}{F1}                                                                    & \multicolumn{3}{c|}{Accuracy}                                                                    & \multicolumn{1}{c|}{AUC}    \\ \cline{3-15} 
\multicolumn{2}{|c|}{}                                & \multicolumn{1}{c|}{AirTag}  & \multicolumn{1}{c|}{Kairos}   & \multicolumn{1}{c|}{Unicorn} & \multicolumn{1}{c|}{AirTag}  & \multicolumn{1}{c|}{Kairos}   & \multicolumn{1}{c|}{Unicorn} & \multicolumn{1}{c|}{AirTag}  & \multicolumn{1}{c|}{Kairos}  & \multicolumn{1}{c|}{Unicorn} & \multicolumn{1}{c|}{AirTag}  & \multicolumn{1}{c|}{Kairos}   & \multicolumn{1}{c|}{Unicorn} & \multicolumn{1}{c|}{Kairos} \\ \hline
\multicolumn{2}{|l|}{E3-CADETS}                       & \multicolumn{1}{c|}{-}       & \multicolumn{1}{r|}{100.00\%} & 98.00\%                      & \multicolumn{1}{c|}{-}       & \multicolumn{1}{r|}{100.00\%} & 100.00\%                     & \multicolumn{1}{c|}{-}       & \multicolumn{1}{c|}{-}       & 99.00\%                      & \multicolumn{1}{c|}{-}       & \multicolumn{1}{r|}{100.00\%} & 99.00\%                      & \multicolumn{1}{c|}{-}      \\ \hline
\multicolumn{2}{|l|}{E3-THEIA}                        & \multicolumn{1}{c|}{-}       & \multicolumn{1}{r|}{100.00\%} & 100.00\%                     & \multicolumn{1}{c|}{-}       & \multicolumn{1}{r|}{100.00\%} & 100.00\%                     & \multicolumn{1}{c|}{-}       & \multicolumn{1}{c|}{-}       & 100.00\%                     & \multicolumn{1}{c|}{-}       & \multicolumn{1}{r|}{100.00\%} & 100.00\%                     & \multicolumn{1}{c|}{-}      \\ \hline
\multicolumn{2}{|l|}{E3-ClearScope}                   & \multicolumn{1}{c|}{-}       & \multicolumn{1}{r|}{100.00\%} & 98.00\%                      & \multicolumn{1}{c|}{-}       & \multicolumn{1}{r|}{100.00\%} & 100.00\%                     & \multicolumn{1}{c|}{-}       & \multicolumn{1}{c|}{-}       & 99.00\%                      & \multicolumn{1}{c|}{-}       & \multicolumn{1}{r|}{100.00\%} & 99.00\%                      & \multicolumn{1}{c|}{-}      \\ \hline
\multicolumn{1}{|l|}{\multirow{2}{*}{C\_15min}} & avg & \multicolumn{1}{r|}{4.04\%}  & \multicolumn{1}{r|}{49.73\%}  & 100.00\%                     & \multicolumn{1}{r|}{70.64\%} & \multicolumn{1}{r|}{97.08\%}  & 94.25\%                      & \multicolumn{1}{r|}{6.17\%}  & \multicolumn{1}{r|}{58.87\%} & 96.52\%                      & \multicolumn{1}{r|}{80.07\%} & \multicolumn{1}{r|}{57.03\%}  & 94.25\%                      & 61.43\%                     \\ \cline{2-15} 
\multicolumn{1}{|l|}{}                          & std & \multicolumn{1}{r|}{11.21\%} & \multicolumn{1}{r|}{35.30\%}  & 0.00\%                       & \multicolumn{1}{r|}{33.71\%} & \multicolumn{1}{r|}{4.25\%}   & 5.53\%                       & \multicolumn{1}{r|}{17.00\%} & \multicolumn{1}{r|}{33.81\%} & 3.63\%                       & \multicolumn{1}{r|}{6.77\%}  & \multicolumn{1}{r|}{29.23\%}  & 5.53\%                       & 8.85\%                      \\ \hline
\multicolumn{1}{|l|}{\multirow{2}{*}{G\_15min}} & avg & \multicolumn{1}{r|}{0.09\%}  & \multicolumn{1}{r|}{43.24\%}  & 99.90\%                      & \multicolumn{1}{r|}{61.24\%} & \multicolumn{1}{r|}{91.39\%}  & 97.38\%                      & \multicolumn{1}{r|}{0.18\%}  & \multicolumn{1}{r|}{53.71\%} & 98.29\%                      & \multicolumn{1}{r|}{80.23\%} & \multicolumn{1}{r|}{57.32\%}  & 97.38\%                      & 67.63\%                     \\ \cline{2-15} 
\multicolumn{1}{|l|}{}                          & std & \multicolumn{1}{r|}{0.13\%}  & \multicolumn{1}{r|}{27.46\%}  & 0.31\%                       & \multicolumn{1}{r|}{27.73\%} & \multicolumn{1}{r|}{12.38\%}  & 4.86\%                       & \multicolumn{1}{r|}{0.26\%}  & \multicolumn{1}{r|}{25.35\%} & 3.39\%                       & \multicolumn{1}{r|}{12.95\%} & \multicolumn{1}{r|}{22.96\%}  & 4.86\%                       & 14.32\%                     \\ \hline
\multicolumn{1}{|l|}{\multirow{2}{*}{M\_15min}} & avg & \multicolumn{1}{r|}{6.41\%}  & \multicolumn{1}{r|}{48.55\%}  & 94.60\%                      & \multicolumn{1}{r|}{62.19\%} & \multicolumn{1}{r|}{82.32\%}  & 78.88\%                      & \multicolumn{1}{r|}{9.47\%}  & \multicolumn{1}{r|}{55.44\%} & 82.00\%                      & \multicolumn{1}{r|}{77.53\%} & \multicolumn{1}{r|}{49.65\%}  & 78.88\%                      & 51.87\%                     \\ \cline{2-15} 
\multicolumn{1}{|l|}{}                          & std & \multicolumn{1}{r|}{14.49\%} & \multicolumn{1}{r|}{27.84\%}  & 3.89\%                       & \multicolumn{1}{r|}{28.93\%} & \multicolumn{1}{r|}{9.86\%}   & 4.96\%                       & \multicolumn{1}{r|}{20.50\%} & \multicolumn{1}{r|}{25.74\%} & 5.44\%                       & \multicolumn{1}{r|}{7.92\%}  & \multicolumn{1}{r|}{20.60\%}  & 4.96\%                       & 7.55\%                      \\ \hline
\multicolumn{1}{|l|}{\multirow{2}{*}{C\_1hour}} & avg & \multicolumn{1}{r|}{2.76\%}  & \multicolumn{1}{r|}{40.64\%}  & 83.55\%                      & \multicolumn{1}{r|}{76.65\%} & \multicolumn{1}{r|}{85.95\%}  & 95.13\%                      & \multicolumn{1}{r|}{4.84\%}  & \multicolumn{1}{r|}{52.83\%} & 83.55\%                      & \multicolumn{1}{r|}{76.26\%} & \multicolumn{1}{r|}{55.27\%}  & 85.20\%                      & 57.80\%                     \\ \cline{2-15} 
\multicolumn{1}{|l|}{}                          & std & \multicolumn{1}{r|}{6.01\%}  & \multicolumn{1}{r|}{24.39\%}  & 11.66\%                      & \multicolumn{1}{r|}{32.45\%} & \multicolumn{1}{r|}{29.63\%}  & 4.22\%                       & \multicolumn{1}{r|}{10.33\%} & \multicolumn{1}{r|}{26.60\%} & 11.66\%                      & \multicolumn{1}{r|}{13.28\%} & \multicolumn{1}{r|}{15.14\%}  & 10.85\%                      & 17.66\%                     \\ \hline
\multicolumn{1}{|l|}{\multirow{2}{*}{G\_1hour}} & avg & \multicolumn{1}{r|}{0.04\%}  & \multicolumn{1}{r|}{38.91\%}  & 85.20\%                      & \multicolumn{1}{r|}{60.49\%} & \multicolumn{1}{r|}{96.47\%}  & 96.25\%                      & \multicolumn{1}{r|}{0.09\%}  & \multicolumn{1}{r|}{53.37\%} & 85.20\%                      & \multicolumn{1}{r|}{77.36\%} & \multicolumn{1}{r|}{45.28\%}  & 87.13\%                      & 56.02\%                     \\ \cline{2-15} 
\multicolumn{1}{|l|}{}                          & std & \multicolumn{1}{r|}{0.05\%}  & \multicolumn{1}{r|}{16.79\%}  & 11.69\%                      & \multicolumn{1}{r|}{28.83\%} & \multicolumn{1}{r|}{6.79\%}   & 5.42\%                       & \multicolumn{1}{r|}{0.10\%}  & \multicolumn{1}{r|}{18.04\%} & 11.69\%                      & \multicolumn{1}{r|}{13.25\%} & \multicolumn{1}{r|}{14.58\%}  & 11.49\%                      & 7.05\%                      \\ \hline
\multicolumn{1}{|l|}{\multirow{2}{*}{M\_1hour}} & avg & \multicolumn{1}{r|}{1.82\%}  & \multicolumn{1}{r|}{33.63\%}  & 87.92\%                      & \multicolumn{1}{r|}{75.68\%} & \multicolumn{1}{r|}{93.09\%}  & 97.90\%                      & \multicolumn{1}{r|}{3.42\%}  & \multicolumn{1}{r|}{47.32\%} & 87.92\%                      & \multicolumn{1}{r|}{78.49\%} & \multicolumn{1}{r|}{46.00\%}  & 90.06\%                      & 58.14\%                     \\ \cline{2-15} 
\multicolumn{1}{|l|}{}                          & std & \multicolumn{1}{r|}{3.03\%}  & \multicolumn{1}{r|}{17.80\%}  & 9.86\%                       & \multicolumn{1}{r|}{17.94\%} & \multicolumn{1}{r|}{8.63\%}   & 4.28\%                       & \multicolumn{1}{r|}{5.66\%}  & \multicolumn{1}{r|}{20.51\%} & 9.86\%                       & \multicolumn{1}{r|}{13.15\%} & \multicolumn{1}{r|}{16.02\%}  & 9.44\%                       & 9.40\%                      \\ \hline
\end{tabular}}
\begin{tablenotes}
        \item C, G, and M denote known, randomly generated, and composite campaigns.
    \end{tablenotes}
    \end{table*}

\subsection{APT Intrusion Detection}\label{sec: intrusion system}
APT intrusion detection is defined as the task of identifying potential APT attacks within an audit log, as established in previous studies~\cite{ding2023airtag,cheng2023kairos,han2020unicorn}. To demonstrate the usefulness of SAGA datasets in supporting APT intrusion detection, we present the detection results using their own evaluation schemes of various methods and then discuss the causes of performance difference, considering the following factors: test granularity, Technique used, attack scenario, and duration.
Table~\ref{tab:6_3_average_intrusion_result} presents the performance of the three intrusion detection systems, AirTag (event-based), Kairos (time window-based) and Unicorn (graph-based) across various SAGA datasets. More details can be found in Tables S5, S6, and S7 of Supplementary Material B.
Performance data for E3-CADETS, E3-THEIA, and E3-ClearScore were copied directly from their respective original papers.
Unicorn, which analyzes the entire provenance graph to detect anomalies, generally performed well in both precision and recall. AirTag, on the other hand, which identifies intrusions by analyzing individual events, struggled with high rates of false positives and false negatives. 
When exposed to SAGA’s broader technique coverage, multi-stage campaign structure, and temporal variation, AirTag, a one-class SVM anomaly detection model built on a customized BERT embedding, struggles to find a stable decision boundary across such diverse contexts.
Kairos was designed with specific assumptions about event sequence timing, which may not align with the more flexible and richly structured scenarios generated by SAGA.
By grouping events within time windows, 
Kairos tended to reduce both false positives and false negatives, resulting in relatively low precision but higher recall.

\noindent \textbf{Test granularity.} 
Test granularity refers to the unit of target tested in a detection method. For example, the text granularity of AirTag is based on individual events. In a test using the APT28 campaign dataset (15 minutes), AirTag generated 301,069 false positives and missed 14,098 malicious events out of 1,209,895 total events. On the other hand, Kairos, which uses time window as its test granularity, incorrectly flagged 42 benign windows and missed just one malicious window out of 67 total windows. 
Table~\ref{tab:6_3_average_intrusion_result} presents the performance of the three detection systems, each using its own scoring method (i.e., a formula to calculate the performance metrics).
This highlights the significant influence of the test granularity and scoring method on the reported results, making direct comparisons between performance data difficult and, at times, not meaningful. Since SAGA provides labels for each event, its datasets are versatile enough to support various test granularity and scoring methods.

\noindent \textbf{Technique impact.}
The generated campaigns (G) feature a greater number and diversity of Techniques compared to the synthetic known campaigns (C). To assess the impact of varying the number and types of Techniques used in different campaigns, we provide details of the audit logs for the known campaigns in Table~\ref{tab:known_campaigns}, and for the generated campaigns in Table S2 of Supplementary Material A. As shown, the audit logs for the generated campaigns contain more and a broader range of Techniques than those in the synthetic known campaigns.
Table~\ref{tab:6_3_average_intrusion_result} presents the average performance of each detection system, showing that the C group consistently outperforms the G group. This decline in performance is directly correlated with the increasing number and diversity of Techniques.

\noindent \textbf{Attack Scenario Impact.}
SAGA is capable of generating synthetic audit logs containing multiple APT campaigns, referred to as composite campaigns (M group), designed to reflect complex real-world scenarios. Details of these composite campaigns, labeled M1 to M10, are provided in Table S7 in the Supplementary Material B. As the complexity of the M group audit logs increases, the detection performance gradually declines (e.g., from C\_15min to M\_15min), as shown in Table~\ref{tab:6_3_average_intrusion_result}.

\noindent \textbf{Duration Impact.}
To demonstrate its ability to generate synthetic audit logs of varying length, 15 minute and 1 hour attack scenarios were generated to detect APT intrusions. An APT attack typically employs a “low-and-slow” approach, maintaining a prolonged presence on the target machine, which makes the intrusion difficult to detect. As shown in Table~\ref{tab:6_3_average_intrusion_result}, the detection performance gradually decreases with longer durations for the same context.

In summary, synthetic audit logs can be tailored in terms of duration, the number and type of embedded APT campaigns, and the types of APT campaigns. As the complexity of these scenarios increases, the performance of detection methods tends to degrade.

\subsection{Evaluation on Technique Hunting}
To demonstrate the practical utility of fine-grained Technique labeling, Technique hunting is the task of identifying potential threats, specifically Techniques, as defined by the MITRE ATT\&CK framework. 
SAGA follows Algorithm \ref{alg:campaign_gen} to generate attack campaigns, where each event is labeled with its corresponding Technique identifier. This structured synthetic audit log allows detection methods to effectively perform Technique hunting.
The approaches are used for Technique hunting; one is SFM and the other is using Sigma rules.
The performance results of Technique hunting for each APT attack campaign are reported in Table~\ref{tab:ttp_hunting} that details are given in Table S9 in the Supplementary Material C. The results indicate significant performance distinctions between SFM and Sigma rules, regardless of the generation method, attack scenario, or duration.
SFM is capable of recognizing attack patterns by learning from synthetic data, whereas Sigma rules are limited to detecting attack patterns that match pre-defined signatures.
This underscores the importance of the generated datasets in facilitating the advancement of learning-based model methodologies.

\begin{table}[tb!]
\caption{The performance of Technique Hunting.}
\label{tab:ttp_hunting}
\centering
\setlength{\tabcolsep}{1mm}{
\begin{tabular}{|l|rr|rr|rr|}
\hline
\multicolumn{1}{|c|}{\multirow{2}{*}{Campaign}} & \multicolumn{2}{c|}{Precision}                          & \multicolumn{2}{c|}{Recall}                             & \multicolumn{2}{c|}{F1}                                 \\ \cline{2-7} 
\multicolumn{1}{|c|}{}                          & \multicolumn{1}{c|}{Sigma}   & \multicolumn{1}{c|}{SFM} & \multicolumn{1}{c|}{Sigma}   & \multicolumn{1}{c|}{SFM} & \multicolumn{1}{c|}{Sigma}   & \multicolumn{1}{c|}{SFM} \\ \hline
C\_15min                                        & \multicolumn{1}{r|}{5.58\%}  & 58.67\%                  & \multicolumn{1}{r|}{8.50\%}  & 81.20\%                  & \multicolumn{1}{r|}{2.34\%}  & 63.86\%                  \\ \hline
G\_15min                                        & \multicolumn{1}{r|}{30.44\%} & 53.65\%                  & \multicolumn{1}{r|}{7.49\%}  & 83.79\%                  & \multicolumn{1}{r|}{9.04\%}  & 58.85\%                  \\ \hline
M\_15min                                        & \multicolumn{1}{r|}{4.61\%}  & 65.27\%                  & \multicolumn{1}{r|}{7.46\%}  & 87.39\%                  & \multicolumn{1}{r|}{2.16\%}  & 68.50\%                  \\ \hline
C\_1hour                                        & \multicolumn{1}{r|}{9.35\%}  & 57.35\%                  & \multicolumn{1}{r|}{11.13\%} & 79.63\%                  & \multicolumn{1}{r|}{2.67\%}  & 62.72\%                  \\ \hline
G\_1hour                                        & \multicolumn{1}{r|}{45.54\%} & 39.98\%                  & \multicolumn{1}{r|}{8.65\%}  & 84.38\%                  & \multicolumn{1}{r|}{11.92\%} & 47.84\%                  \\ \hline
M\_1hour                                        & \multicolumn{1}{r|}{4.62\%}  & 47.43\%                  & \multicolumn{1}{r|}{7.46\%}  & 87.73\%                  & \multicolumn{1}{r|}{2.20\%}  & 56.27\%                  \\ \hline
C\_1day                                         & \multicolumn{1}{r|}{0.38\%}  & 5.04\%                   & \multicolumn{1}{r|}{8.63\%}  & 77.22\%                  & \multicolumn{1}{r|}{0.58\%}  & 9.21\%                   \\ \hline
G\_1day                                         & \multicolumn{1}{r|}{6.00\%}  & 2.39\%                   & \multicolumn{1}{r|}{10.20\%} & 83.71\%                  & \multicolumn{1}{r|}{6.28\%}  & 4.51\%                   \\ \hline
M\_1day                                         & \multicolumn{1}{r|}{0.83\%}  & 6.87\%                   & \multicolumn{1}{r|}{7.05\%}  & 88.44\%                  & \multicolumn{1}{r|}{1.03\%}  & 11.88\%                  \\ \hline
\end{tabular}}
\end{table}

However, both methods showed significant performance fluctuations across various attack scenarios. For example, the average standard deviations of precision during the 15-min scenarios were 33.92\% and 32.25\% for SFM and Sigma rules, respectively. 
Several factors contributed to this variability. Firstly, there was a high incidence of false positives associated with specific attack behaviors. Secondly, the substantial imbalance between the number of malicious events and the overwhelming prevalence of benign activities skewed the evaluation of attack behaviors. Lastly, benign activities and malicious events were intertwined and exhibited different patterns in various datasets, complicating the recognition of attack patterns.

For instance, recognizing T1105 Ingress Tool Transfer (ITT3) requires identifying the following sequence of audit events within interleaved benign event sequences: \textless{}\textit{browser.exe}, \textit{TCP connect}, \textit{XXX.com}\textgreater{}, \textless{}\textit{browser.exe}, \textit{FileCreate}, \textit{payload.exe}\textgreater{}, and \textless{}\textit{xxx}, \textit{Process Create}, \textit{payload.exe}\textgreater{}. 
This sequence indicates that a malicious payload was downloaded from an external C2 server and executed on a compromised machine. These malicious events are causally correlated and in audit log, they were intertwined with benign and background events, making the recognition of this Technique particularly challenging.

On the other hand, while Sigma rules perform efficiently, they are based on contributions from human experts and may not encompass all attack behaviors, such as T1055.002 Portable Executable Injection (PEI) and T1491 Defacement (DF2) identified in this empirical study. As a result, certain malicious behaviors went undetected in the APT campaigns C-6, C-8, G-11, and G-13.
These findings address the limitations of manually crafted, signature-based detection methods and highlight the importance of datasets like those generated by SAGA, which provide comprehensive logs from red-team emulation campaigns. 
Such datasets support the development and evaluation of both rule-based detection systems and learning-based models in a realistic environment.

\begin{figure}[tb!]
    \centering
    \includegraphics[width=1\columnwidth]{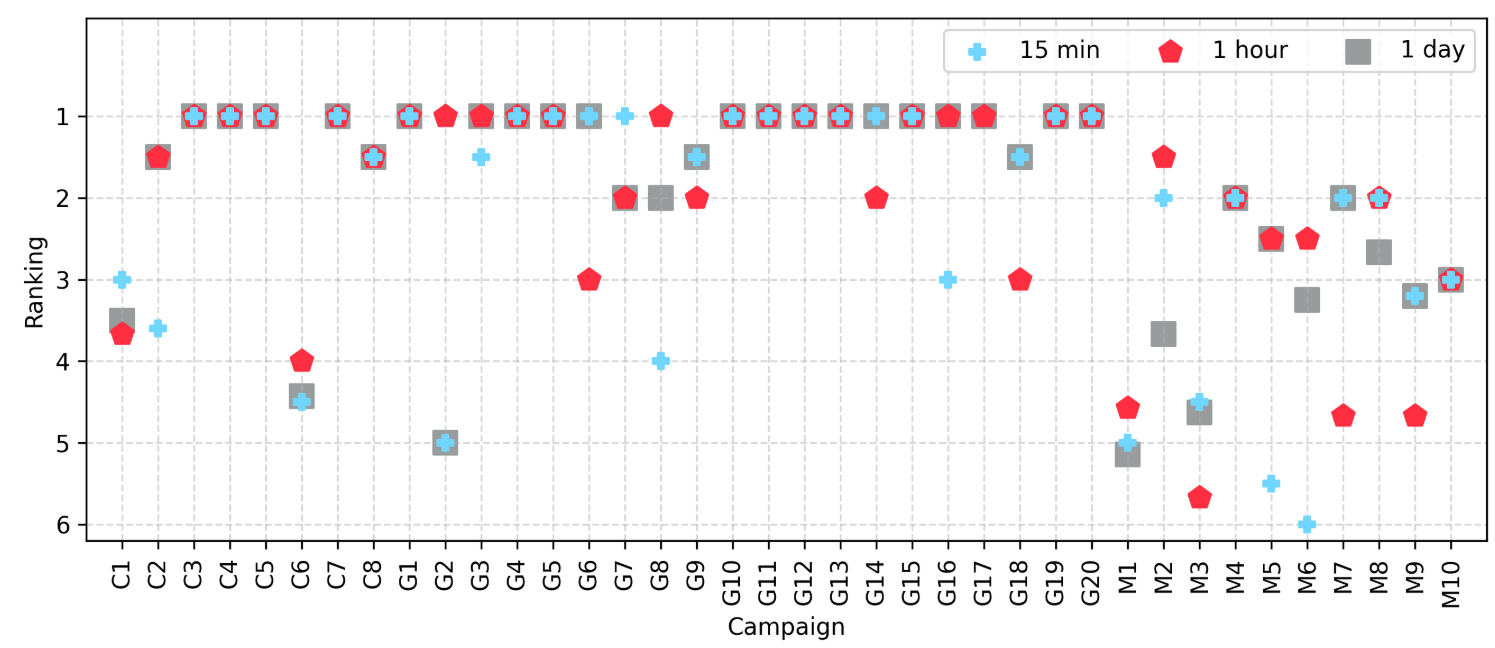}
    \caption{The top $k$ ranking results of APT campaign detection by SFM. }
    \label{fig:Attack_Campaign}
\end{figure}

\subsection{APT Campaign Detection Evaluation}
APT campaign detection involves identifying a potential APT campaign embedded in an audit log, which is characterized by a series of coordinated attack patterns. In line with the MITRE ATT\&CK framework, an APT campaign can be described as a sequence of Techniques. SFM leverages this structure to detect the most likely APT campaign by comparing the discovered Techniques ($G_q$) against a pool of known APT campaigns ($G_c$). In this empirical study, $G_c$ includes 8 known campaigns and 30 randomly generated APT campaigns.
Figure~\ref{fig:Attack_Campaign} illustrates the ranking of each SFM-identified campaign over three different durations, 15 minutes, 1 hour, and 1 day.
SFM successfully identified most campaigns within the top 3 rankings across all three time durations: 74\% for 15 minutes, 84\% for 1 hour, and 79\% for 1 day. 
Notably, 50\% of the campaigns were ranked first in both 1 hour and 1 day scenarios. 
One reason for this success could be that the temporal causal relationships is established among the Techniques discovered by SFM, which effectively distinguishes the most probable campaign from others. 
Furthermore, SFM demonstrated resilience to discrepancies between the discovered ability graph $G_q$ and the campaign Techniques $G_c$.
However, the M group, consisting of multiple campaigns in a single scenario, posed a greater challenge compared to the groups of single campaigns. In addition, the results further suggest that as the duration of audit log grows, the complexity of identifying APT campaigns also increases.
In conclusion, this evaluation highlights the potential of SAGA in effectively supporting the detection of APT campaigns.

\subsection{A Case Study of Unseen APT Campaign}
In this section, we investigate the effectiveness of SAGA synthetic audit log datasets in training an APT campaign model to be applied to previously unseen APT campaigns, even when the campaign audit logs follow a different format and organization.
Specifically, SFM was trained using SAGA datasets and tested on the DARPA TC E3 Fivedirection dataset to detect attack behaviors (i.e., Techniques) within the logs.

\begin{figure*}[!htb]
    \centering
    \includegraphics[width=0.8\textwidth]{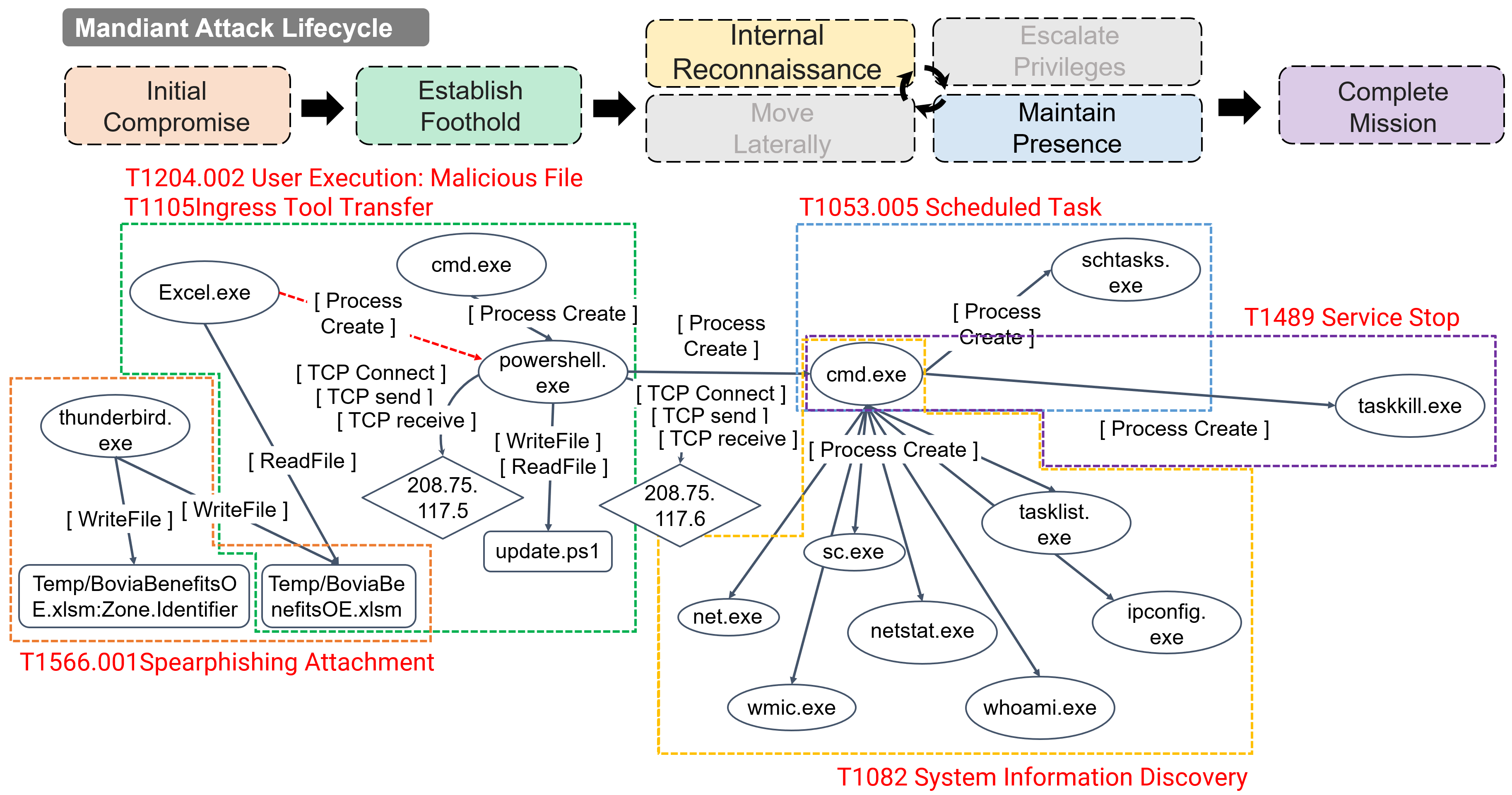}
    \caption{The DARPA TC E3 Fivedirection attack graph}
    \label{fig:darpa_e3}
\end{figure*}

The E3 Fivedirection dataset includes a complex attack scenario (4.4  Phishing Email with Excel Macro~\cite{darpa2data}). As illustrated in Figure~\ref{fig:darpa_e3}, the attacker used a phishing attachment to lure the victim into downloading and executing malware that triggered subsequent malicious activities, including establishing a remote command shell and gathering various system information. From the detection results, it shows that SFM was able to successfully identify part of the attack sequence. For example, during the initial compromise phase, the SFM successfully flagged the system events corresponding to the victim downloading the phishing attachment, \textless{}\textit{thunderbird.exe}, \textit{WriteFile}, \textit{BoviaBenefitsOE.xlsm}\textgreater{}, as T1566.001 Spearphishing Attachment. Additionally, during the Establish Foothold phase, the system events corresponding to the execution of the phishing attachment, \textless{}\textit{Excel.exe}, \textit{CreateFile}, \textit{BoviaBenefitsOE.xlsm}\textgreater{}, were also correctly flagged as T1204.002 User Execution: Malicious File.

However, some attack behaviors were not correctly identified by SFM. For example, during the Establish Foothold phase, when the attacker used PowerShell to download a malicious payload (\textit{update.ps1}) from a C2 server, SFM successfully marked the system events related to the download activity, \textless{}\textit{powershell.exe}, \textit{CreateFile}, \textit{update.ps1}\textgreater{}, as T1105 Ingress Tool Transfer, but it failed to include the system events, PowerShell connection to the C2 server, \textless{}\textit{powershell.exe}, \textit{TCP connect}, \textit{208.75.117.5}\textgreater{} in T1105. This limitation could be attributed to gaps in the SAGA dataset’s representation of certain attack behaviors, or SFM's difficulty in detecting subtle variations in event sequences.

However, these results highlight the practical value of the SAGA dataset in supporting practical cybersecurity applications and providing a solid foundation for future improvements to dataset generation strategies. They also underscore the need for further improvements in APT campaign detection models to enhance their capability and accuracy.

\section{Related Work}
\noindent\textbf{Existing Benchmarks for Host-based Intrusion Detection.}
As noted in the survey by~\cite{zipperle2022provenance}, the majority of studies on host-based intrusion detection have relied on self-constructed attack scenarios for evaluation. 
Some works have utilized existing public datasets such as CERT’s dataset~\cite{glasser2013bridging},  LANL’s dataset~\cite{kent2015comprehensive}, StreamSpot~\cite{manzoor2016steampot} ADFA~\cite{creech2014ADFA}, and DARPA Transparent Computing (TC)\cite{darpa2tc}, as well as OpTC~\cite{darpa2optc}.
CERT's dataset provides synthetic background data alongside malicious operations, such as logon and email activities, carried out by multiple threat actors across five distinct attack scenarios. 
LANL's dataset consists of 58 consecutive days of event data collected from LANL's corporate computer network, including logon events, process events, DNS lookups, network flow events, and a set of well-defined red teaming events representing malicious activities. 
StreamSpot includes one attack activity—a drive-by download initiated by visiting a malicious URL that gains root access to the visiting host—along with five benign activities forming three scenarios.
However, these datasets primarily focus on event action types (i.e., operations) and timestamps, lacking critical details regarding the underlying system entities involved in the audit events. This absence limits their ability to clearly describe the full context of each event, which is essential for comprehensive attack pattern analysis and detection.

ADFA comprises three datasets: ADFA-LD (Linux), ADFA-WD (Windows), and ADFA-WD-SAA (Windows for Stealth Attacks).
Among these, ADFA-WD is the most relevant to SAGA. However, it only contains DLL traces of processes collected by Procmon and includes twelve zero-day attacks executed on a host.

DARPA's TC program included five engagements and the OpTC dataset. The E3 and E5 engagements, featuring two and eight multi-stage APTs respectively, are widely used for intrusion detection, alongside benign background activity. The OpTC dataset, created to assess scalability, simulated a three-day APT on a 1,000-host test network during a two-week experiment.
However, these datasets lack thorough documentation and labels, which limits their usefulness for precisely evaluating attack behaviors and audit events.

\noindent\textbf{Synthetic APT Dataset Generation.}
While there are some publicly available attack datasets, few research focus on constructing datasets for simulating multi-stage APT attacks~\cite{myneni2020dapt, berady2022pwnjutsu, myneni2023unraveled, LADEMU}.
To steal organization data, 
DAPT2020~\cite{myneni2020dapt} imitated the exploited web vulnerability in a production network, and Unraveled~\cite{myneni2023unraveled} created skeletons of user profiles to emulate user behavior based on predefined job pipelines.
These datasets encompass network traffic and multi-host log information, covering limited attacks (i.e., 10 and 15) within attack stages over a fixed collection duration (i.e., 1 week or 6 weeks), 
In contrast, our generated APT dataset includes and labels 7 attack stages, 80 techniques, and 169 abilities, offering \textit{configurable} attack patterns and collection duration on single-host audit events.
This flexibility allows for more detailed and customizable simulation of attack behaviors. 

PWNJUTSU~\cite{berady2022pwnjutsu} and LADEMU~\cite{LADEMU} considered adversaries' TTPs, defining the operational semantics of Techniques as specifications for procedures. PWNJUTSU utilized a semi-automated approach where each attack step was crafted by professional human Red Teamers, providing a highly skilled yet partially manual campaign construction process. In contrast, LADEMU offered a proof-of-concept (PoC) implementation using the red-team emulator Caldera and demonstrated the use of a single attack scenario, APT29.
Our study introduces a workflow of campaign generation with event logs, enabling fully automated APT simulations based on TTPs specified in the MITRE ATT\&CK framework. 
This approach eliminates the need for manual intervention, ensuring more consistent and scalable attack simulations.
Table~\ref{tab:comparison} contrasts SAGA with previous framework and highlights SAGA's unique advantages, particularly its configurability (in terms of duration, techniques, and campaign design), fine-grained labeling, and high degree of automation.

\noindent \textbf{Host-based Provenance analysis.}
Learning-based provenance analysis methods have shown notable improvements in the detection of intrusions and have accelerated the investigative process of various attacks~\cite{zipperle2022provenance, han2020unicorn, alsaheel2021atlas, zengy2022shadewatcher, ding2023airtag, cheng2023kairos, jia2023magic, sharif2024drsec}.

These learning-based models employ a range of representation learning techniques to encode provenance graphs, enabling effective anomaly detection. 
Examples include graph representation methods used in Unicorn~\cite{han2020unicorn}, Kairos~\cite{cheng2023kairos}, and MAGIC~\cite{jia2023magic}; translation-based embedding techniques in WATSON~\cite{zeng2021watson} and ShadeWatcher~\cite{zengy2022shadewatcher}; and language models in AirTag~\cite{ding2023airtag} and DrSec~\cite{sharif2024drsec}.
 Our research demonstrates that applying event-based~\cite{ding2023airtag} and graph-based intrusion detection~\cite{han2020unicorn, cheng2023kairos} to our synthesized campaign datasets results in effective detection of intrusion events.

\noindent
\textbf{Technique Hunting.}
HOLMES~\cite{milajerdi2019holmes} generates detection signals by matching 16 predefined attack patterns and constructing a high-level scenario graph embedded with TTPs to indicate APT campaign activities. 
RapSheet~\cite{hassan2020tactical} creates a tactical provenance graph using 67 rules within a commercial EDR tool, aligning it with MITRE ATT\&CK. 
KRYSTAL~\cite{kurniawan2022krystal} leverages Sigma rules~\cite{sigma2} to detect known attack patterns, converting them into SPARQL queries for threat detection. 
Although these rule-based approaches are effective, their manual creation and upkeep are resource-intensive. 
Our study addresses this limitation by introducing a flexible approach to APT dataset generation using a red-team emulator for audit event collection, enabling a learning-based methodology. Additionally, we propose a straightforward method (SFM) for identifying attack patterns, providing a baseline for future research.

\begin{table}[h!]
\caption{Comparisons with Synthetic APT Dataset Generation.}
\label{tab:comparison}
\resizebox{0.5\textwidth}{!}{
\begin{tabular}{|c|c|c|c|c|c|}
\hline
\begin{tabular}[c]{@{}c@{}}Log\\ Generation\end{tabular} & \begin{tabular}[c]{@{}c@{}}DAPT2020\\ \cite{myneni2020dapt}\end{tabular} & \begin{tabular}[c]{@{}c@{}}PWNJUTSU\\ \cite{berady2022pwnjutsu}\end{tabular} & \begin{tabular}[c]{@{}c@{}}Unraveled\\ \cite{myneni2023unraveled}\end{tabular} & \begin{tabular}[c]{@{}c@{}}LADEMU\\ \cite{LADEMU}\end{tabular} & SAGA \\ \hline
Duration & 3 months & - & 6 weeks & 1 hour & Configurable \\ \hline
\# Attackers & 1 & 22 & 3 & 1 & Configurable \\ \hline
Attack Stage & 4 & 5 & 5 & 4 & 6 \\ \hline
\# Attacks & 16 & 13 & 15 & 79 & \begin{tabular}[c]{@{}c@{}}197\\ (extendable)\end{tabular} \\ \hline
Label & \begin{tabular}[c]{@{}c@{}}Stage,\\ Attack \\ Behavior\end{tabular} & \begin{tabular}[c]{@{}c@{}}Stage, \\ Technique\end{tabular} & \begin{tabular}[c]{@{}c@{}}Activity, \\ Stage, \\ Defender \\ Response, \\ Signature\end{tabular} & \begin{tabular}[c]{@{}c@{}}Tactic, \\ Technique\end{tabular} & \begin{tabular}[c]{@{}c@{}}Stage, \\ Technique, \\ Ability, \\ Campaign\end{tabular} \\ \hline
\begin{tabular}[c]{@{}c@{}}Alignment \\ to \\ MITRE \\ ATT\&CK\end{tabular} & N & Y & Y & Y & Y \\ \hline
\begin{tabular}[c]{@{}c@{}}Threat \\ Model\end{tabular} & \begin{tabular}[c]{@{}c@{}}Human\\ Experts\end{tabular} & \begin{tabular}[c]{@{}c@{}}Human\\ Experts\end{tabular} & \begin{tabular}[c]{@{}c@{}}Employee \\ Behavior \\ Generation\end{tabular} & \begin{tabular}[c]{@{}c@{}}Red team \\ emulator \\ (Caldera)\end{tabular} & \begin{tabular}[c]{@{}c@{}}Red team \\ emulator \\ (Caldera)\end{tabular} \\ \hline
Automation & \halfcirc & \halfcirc & \halfcirc & \begin{tabular}[c]{@{}c@{}}\fullcirc\\ (PoC) \end{tabular}
& \fullcirc \\ \hline
\end{tabular}
}
\end{table}

\section{Discussion and Conclusion}
This study presents a generative APT campaign framework, SAGA, designed to synthesize synthetic audit logs for nearly any configuration. These logs support the training of various machine learning-based APT-related detection machine learning models and facilitate the evaluation of various APT-related detection methods. To demonstrate the usefulness of SAGA’s synthetic datasets, a collection of empirical studies were conducted across multiple established baselines for APT intrusion detection, Technique hunting, and APT campaign detection. 
In light of the observation of artifact leakage, distributional shift, benign syntheticity, and attack covertness in the synthetic datasets~\cite{liu2025we}, we emphasize that the SAGA-generated datasets are not limited to single-campaign training or evaluation. They can be flexibly composed to include multiple campaigns, time spans, and benign activities, allowing researchers to design training and testing splits that better reflect real-world conditions and capture diverse, multi-stage attack behaviors. Furthermore, SAGA enables researchers to explore distributional shifts by generating stage-specific audit logs, supporting the training and evaluation of models on early-stage versus late-stage attack behaviors.

The authenticity of SAGA-generated logs is grounded in two factors: the faithful implementation of attack techniques and the construction of realistic attack sequences. 
SAGA relies on a red-team emulator (i.e., Caldera) to execute concrete implementations of ATT\&CK techniques. The more faithfully the emulator reproduces adversary behaviors, the more realistic the resulting logs.
For sequence construction, SAGA supports both expert-instructed campaigns, emulating real-world APT scenarios as defined by domain experts (similar to manual red-teaming), and sampled sequences, assembled from statistically plausible and logically coherent technique transitions. This design enables SAGA to generate scalable and scenario-driven audit logs, addressing the broader need for labeled and scenario-driven log data for learning-based APT detection research.

The quality of SAGA datasets depends on access to extensive, high-quality campaign knowledge and diverse Technique implementations. 
The current implementation of SAGA uses 169 abilities and 80 Techniques, with ongoing efforts to expand these numbers to enhance the usefulness of SAGA datasets and cover more sophisticated APT campaigns. 
While the ATT\&CK framework is continuously evolving, SAGA is designed to be extensible. New techniques can be integrated by either updating Caldera with additional abilities as they become available, or manually constructing new attack templates through expert-guided red-team emulation. 
Furthermore, we also recognize the potential of advanced tools such as large language models (LLMs) used in~\cite{singer2025feasibility} to assist in the automatic attack generation -- a promising direction for future research.
Given the inherent differences between synthetic and real-world data, we caution that experimental results obtained using SAGA should be interpreted carefully, as synthetic logs may not present in real environments. As highlighted in~\cite{liu2025we}, careful evaluation is necessary to avoid overestimating generalization capabilities due to unintended signals in synthetic datasets.
In this work, the Windows operating system is used as the target environment; however, the concept can be readily applied to other platforms as well.

\bibliographystyle{IEEEtran}
\bibliography{IEEEabrv, ref.bib}
\vspace{-20 pt}

\begin{IEEEbiography}[{\includegraphics[width=1in,height=1.25in,clip,keepaspectratio]{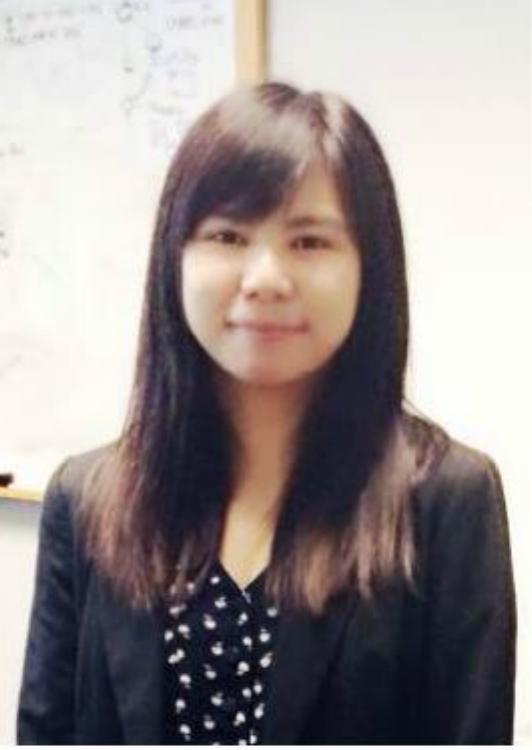}}]{Yi-Ting Huang}
is an assistant professor at National Taiwan University of Science and Technology.
She received the Ph.D.\ degree in Information Management from National Taiwan University in 2015.
She was a postdoctoral fellow at Academia Sinica.
Her primary research interests include malware analysis, cyber threat intelligence analysis, deep learning, and natural language processing.
\end{IEEEbiography}

\begin{IEEEbiography}[{\includegraphics[width=1in,height=1.25in,clip,keepaspectratio]{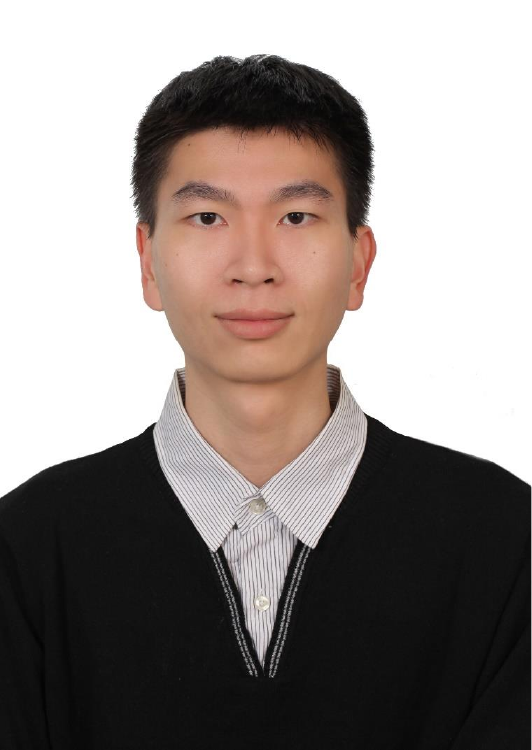}}]{Ying-Ren Guo}
is a research assistant in Academia Sinica. His research interests include malware reverse engineering and analysis, and cybersecurity.
\end{IEEEbiography}

\begin{IEEEbiography}[{\includegraphics[width=1in,height=1.25in,clip,keepaspectratio]{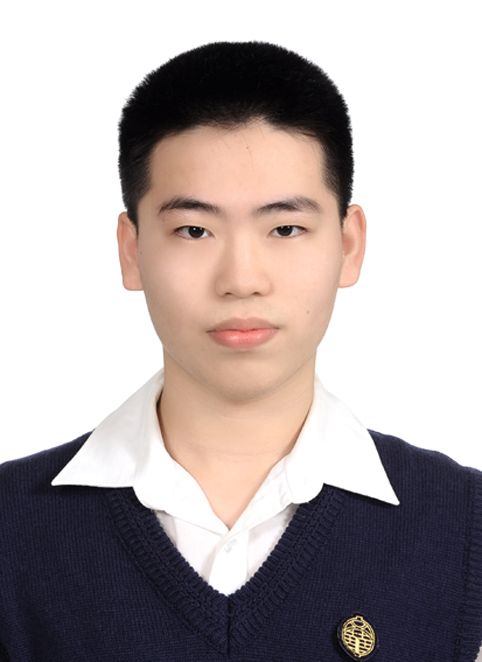}}]{Yu-Sheng Yang}
is a Master's student in Computer Science at National Taiwan University. His research focuses on large language models (LLMs) and computer vision.
\end{IEEEbiography}

\begin{IEEEbiography}[{\includegraphics[width=1in,height=1.25in,clip,keepaspectratio]{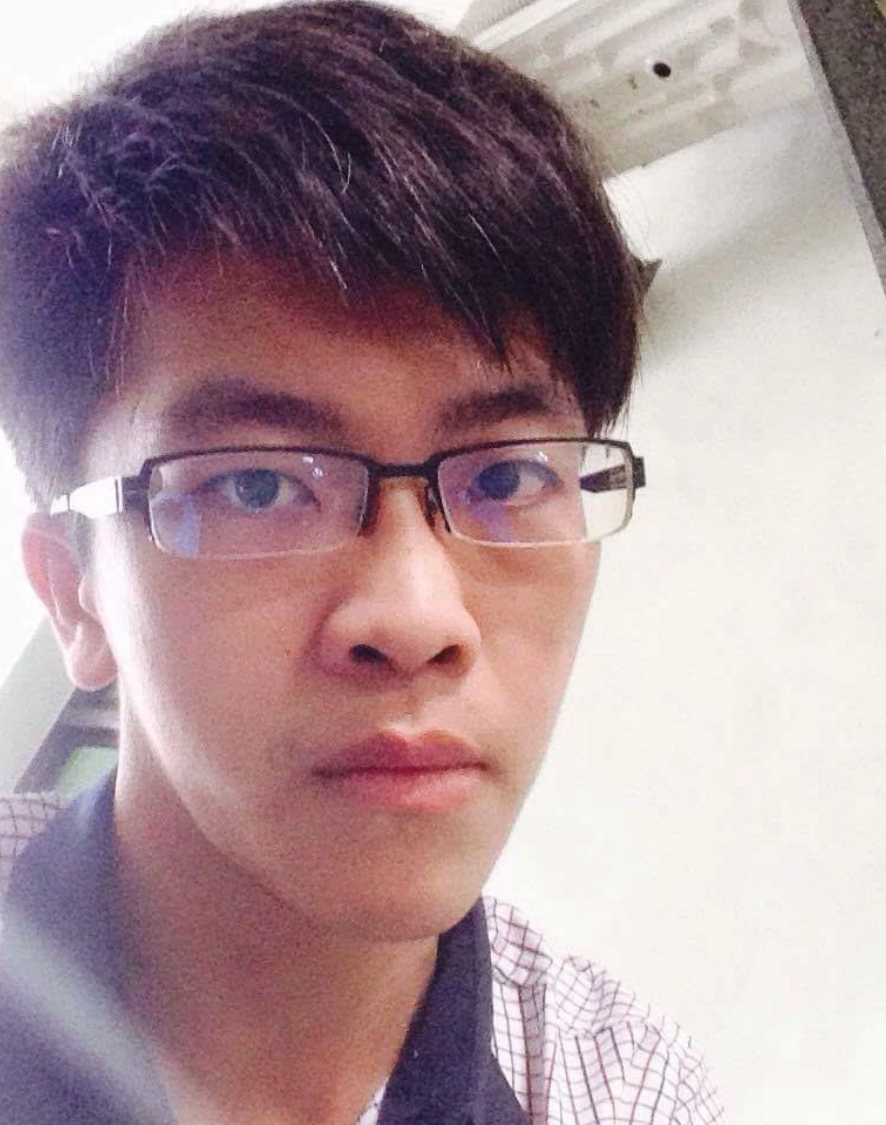}}]{Guo-wei Wong}
received the master degree in thermal and fluid science from the National Yunlin University of Science and Technology in 2017. He is currently working towards his Ph.D. degree of CSIE at the National Taiwan University. His primary research interests include deep learning in fluid dynamics and for complicated applications, including malware analysis and PM2.5 prediction.
\end{IEEEbiography}

\begin{IEEEbiography}[{\includegraphics[width=1in,height=1.25in,clip,keepaspectratio]{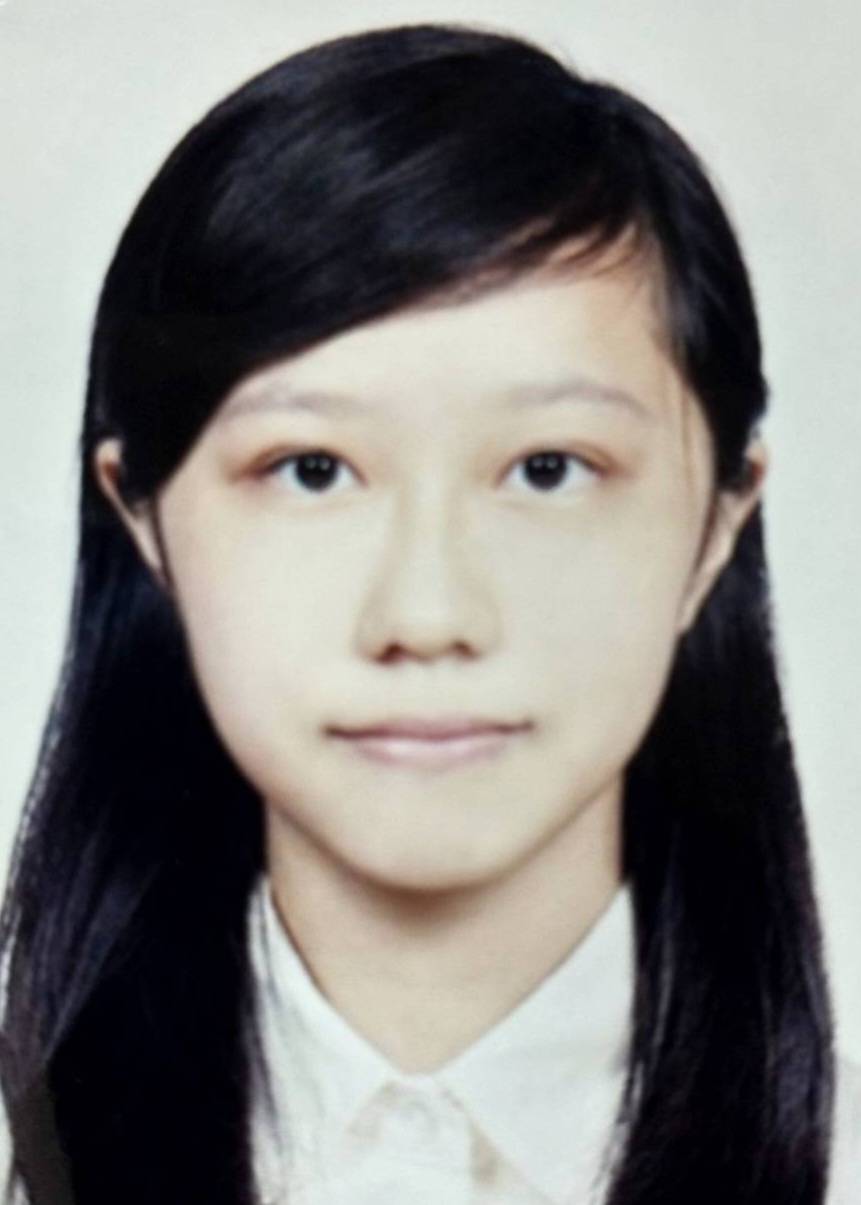}}]{Yu-Zih Jheng} 
is a Master's student in Information Management at National Taiwan University. Her research interests include static and dynamic malware analysis, red teaming, and penetration testing.
\end{IEEEbiography}

\begin{IEEEbiography}[{\includegraphics[width=1in,height=1.25in,clip,keepaspectratio]{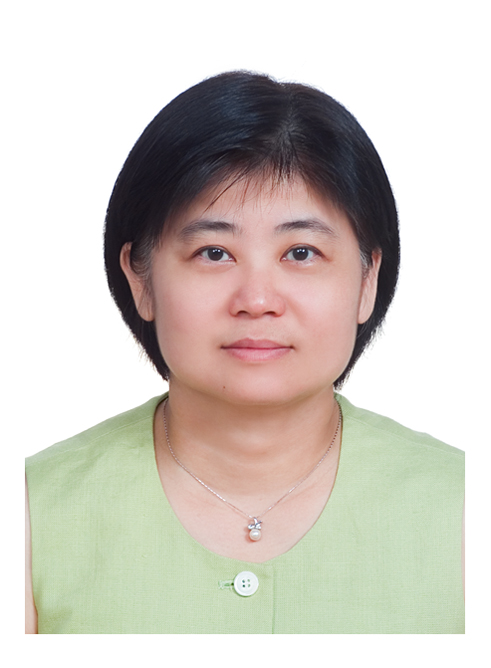}}]{Yeali S. Sun}
received her B.S.\ from the Computer Science and Information Engineering department of National Taiwan University, and the M.S.\ and Ph.D.\ degrees in Computer Science from the University of California, Los Angeles (UCLA). From 1988 to 1993, she was with Bell Communications Research Inc. In August 1993, she joined National Taiwan University and is currently a professor of the Department of Information Management. Her research interests are in the areas of Internet security and forensics, Quality of Service (QoS), cloud computing and services, and performance modeling and evaluation.
\end{IEEEbiography}

\begin{IEEEbiography}[{\includegraphics[width=1in,height=1.25in,clip,keepaspectratio]{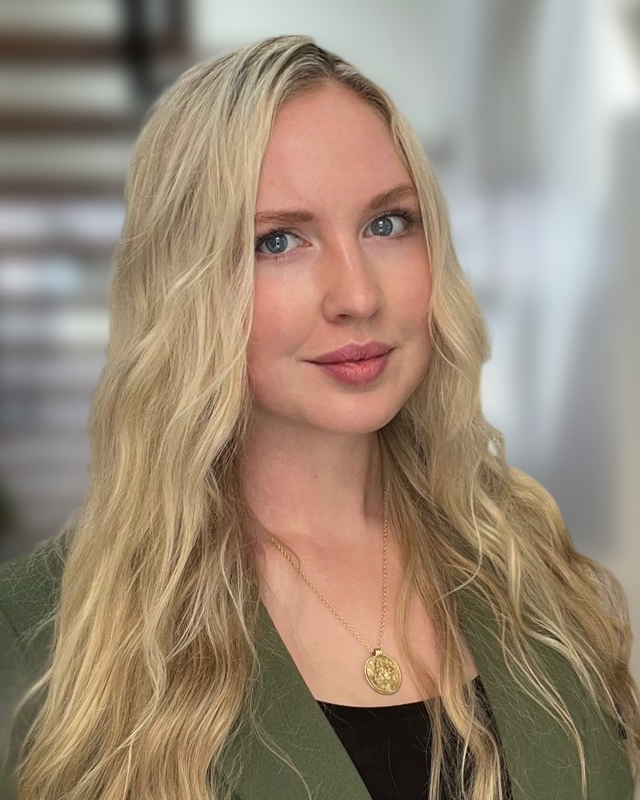}}]{Jessemyn Modini}
is an IT leader and cyber security researcher. Jess holds 5 Master’s qualifications in Cybersecurity Operations, Systems Engineering, Space Operations, Project Management and International Relations (Security). Jess is currently undertaking a Doctorate in Cyber Security, focusing on applications of epidemiology to cyber security (Cyber Epidemiology) for threat modelling and cyber herd immunity at scale. Jess is currently a Senior Technologist at Amazon Web Services where she drives global technology strategy. Prior to AWS, Jess was the Australian Cyber Security Centre (ACSC) Engineering Manager. Jess is passionate about security, global cyber threat intelligence sharing and multidisciplinary approaches to resilience.  
\end{IEEEbiography}

\begin{IEEEbiography}[{\includegraphics[width=1in,height=1.25in,clip,keepaspectratio]{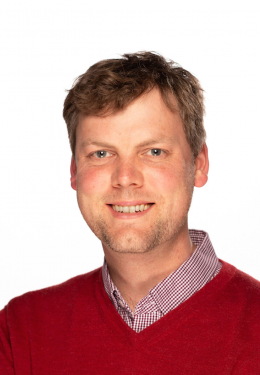}}]{Timothy Lynar} 
is a Senior Lecturer at UNSW Canberra, he received his Ph.D. degree from the University of Newcastle, Australia, in 2011.
He worked at IBM research for 7.5 years where he was a Research Staff member and Master inventor, before joining UNSW in 2019. 
Tim's primary research focus is the application of machine learning to Cyber security. Tim has a background in simulation, modelling and distributed computing including cloud and IoT systems. 
\end{IEEEbiography}

\begin{IEEEbiography}[{\includegraphics[width=1in,height=1.25in,clip,keepaspectratio]{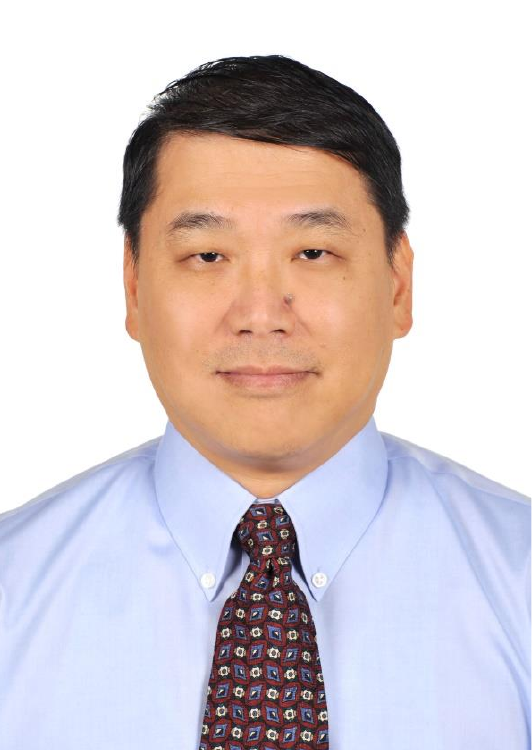}}]{Meng Chang Chen}
received his Ph.D. degree in Computer Science from the University of California, Los Angeles. He was with AT\&T Bell Labs and is now a Research Fellow/Professor of Research Center for Information Technology Innovation, Academia Sinica, Taiwan. His current research interests include computer and network security, wireless network, deep learning for complicated applications, data and knowledge engineering.
\end{IEEEbiography}

\end{CJK*}
\end{document}